\documentclass[aps,twocolumn,epsf,floats,pre,superscriptaddress,nofootinbib]{revtex4-1}

\usepackage{graphicx}
\usepackage{amsmath}
\usepackage{microtype}
\usepackage{tikz}
\usepackage[compat=1.1.0]{tikz-feynman}
\usepackage{commath}
\usepackage{enumitem}
\usepackage{amsfonts}
\usepackage{hyperref}
\usepackage{mathtools}
\usepackage{scrtime}
\usepackage[dont-mess-around]{fnpct}
\usepackage{import}
\usepackage{float}
\usepackage{etoolbox}
\usepackage{placeins}

\newcommand{\MM}{\boldsymbol{M}}
\newcommand{\xxi}{\boldsymbol{\xi}}
\newcommand{\ppsi}{\boldsymbol{\Psi}}
\newcommand{\beq}{\begin{equation}}
\newcommand{\eeq}{\end{equation}}

\begin{document}

\title{Discrete Laplacian Thermostat for Spin Systems with Conserved Dynamics}

\author{Andrea Cavagna}
\affiliation{Istituto Sistemi Complessi (ISC-CNR), Via dei Taurini 19, 00185, Rome, Italy}
\affiliation{Dipartimento di Fisica, Sapienza Universit\`a di Roma, P.le Aldo Moro 2, 00185, Rome, Italy}
    \affiliation{INFN, Unit\`a di Roma 1, 00185 Rome, Italy}
    
\author{Javier Crist\'i­n}
\affiliation{Istituto Sistemi Complessi (ISC-CNR), Via dei Taurini 19, 00185, Rome, Italy}
\affiliation{Dipartimento di Fisica, Sapienza Universit\`a di Roma, P.le Aldo Moro 2, 00185, Rome, Italy}
 
 \author{Irene Giardina}
\affiliation{Dipartimento di Fisica, Sapienza Universit\`a di Roma, P.le Aldo Moro 2, 00185, Rome, Italy}
\affiliation{Istituto Sistemi Complessi (ISC-CNR), Via dei Taurini 19, 00185, Rome, Italy}
    \affiliation{INFN, Unit\`a di Roma 1, 00185 Rome, Italy}
    
\author{Mario Veca}
\affiliation{Dipartimento di Fisica, Sapienza Universit\`a di Roma, P.le Aldo Moro 2, 00185, Rome, Italy}


\begin{abstract}
A well-established numerical technique to study the dynamics of spin systems in which symmetries and conservation laws play an important role is to microcanonically integrate their reversible equations of motion, obtaining thermalization through initial conditions drawn with the canonical distribution. In order to achieve a more realistic relaxation of the magnetic energy, numerically expensive methods that explicitly couple the spins to the underlying lattice are normally employed. Here we introduce a new stochastic conservative thermostat that relaxes the magnetic energy while preserving the constant of motions, thus turning microcanonical spin dynamics into a conservative canonical dynamics, without actually simulating the lattice. We test the thermostat on the Heisenberg antiferromagnet in $d=3$ and show that the method reproduces the exact values of the static and dynamic critical exponents, while in the low temperature phase it yields the correct spin wave phenomenology.  Finally, we demonstrate that the relaxation coefficient of the new thermostat is \textcolor{black}{quantitatively} connected to the microscopic parameters of the spin-lattice coupling.
\end{abstract}

\maketitle

\section{Introduction} \label{Intro}
The impact of symmetries and conservation laws on the dynamics of physical systems cannot be overstated, and spin systems are no exception. Not only conservation laws can change the low-temperature dispersion relations, but they can also radically change the dynamical critical exponents \cite{HH}. The most effective method to numerically study spin systems with symmetries and conservation laws is to microcanonically integrate the reversible equations of motion \cite{windsor1967spin, lurie1974computer}, a technique called Spin Dynamics (SD) by Landau and coworkers, who advanced it very significantly \cite{landau1979computer, gerling1983static, Landau1994, Landau1996, Landau1999_2, Landau2000, Landau2003}. 
As the energy is conserved, in order to thermalize the system SD draws the initial conditions from a canonical ensemble at temperature $T$ using Montecarlo. Although SD provides excellent results, one can raise an issue, which is both conceptual and practical.
Consider a microcanonical simulation of a particles system, as in standard Molecular Dynamics (MD); despite the inevitable simplifications, one can argue that MD is conceptually the same as the actual physical dynamics of an isolated system. On the other hand, microcanonical SD has not quite the same conceptual standing as microcanonical MD: in an actual spin system, where the spins belong to the atoms of an underlying lattice, thermal relaxation occurs mostly through the exchange of energy (but not of magnetization) between the spins and the nuclei; by excluding the lattice from the simulation, and including the temperature only through the initial conditions, SD takes a (clever) shortcut, which has however no actual physical counterpart, as in most physical systems it is quite hard to isolate the spins from the lattice. While within MD one can consider a subsystem A as the heat bath acting upon an adjacent subsystem B, in most spin systems the heat bath is provided by the lattice, which is instead absent in SD. 
This issue has also practical implications; for example, if we want to change the temperature during a spin simulation or if we want to study the effects of a quench, SD has a problem, as $T$ is fixed once and for all at the beginning of the simulation by the initial conditions.
It was precisely to deal with this problem that Tauber and Nandi devised an interesting hybrid method, in which microcanonical SD is alternated with canonical  Kawasaki Montecarlo (KM) at temperature $T$, giving rise to a SD-KM-SD-KM-$\dots$ dynamical sequence \cite{Tauber2019,Tauber2020}. Notice that although KM is conservative, it is not a {\it true} dynamics, as conservation is achieved by swap moves, rather than being dynamically generated by the symmetry through Noether's theorem; which is precisely why KM needs to take turns with SD in the method of \cite{Tauber2019,Tauber2020}. 

A more fundamental approach is to explicitly take into consideration the interaction between spins and lattice by running in parallel a spin dynamics and a molecular dynamics simulation, an approach that we will call SD+MD \cite{Phonon_I,Phonon_II}. Even though in this microcanonical dynamics the total energy is conserved, there is energy exchange between spin and lattice, so that the magnetic energy relaxes. SD+MD is the most complete and realistic numerical method to simulate magnetic systems, but it is also significantly more expensive than SD from a computational point of view, as it needs to update the positions and the momenta of the nuclei, in addition to the spins.  It would be useful to have a method as simple and economic as SD, but which includes the relaxational effects of the spin-lattice coupling. \textcolor{black}{We note that such methods exist for the case where interactions do {\it not} conserve the total magnetization. The Landau-Lifshitz-Gilbert equation \cite{gilbert2004} and other Langevin-type equations \cite{evans2014atomistic,evans2012stochastic} relax the spin energy through multiple dissipative terms, violating the conservation law. However, analogous mechanisms for the conservative case, where the total magnetization is constant, are still lacking.} Here, we fill that gap and present a new stochastic thermostat that turns microcanonical spin dynamics into a conservative canonical dynamics: the novel numerical method updates only the spins, hence having a low computational cost -- similar to SD -- and yet it thermalizes the magnetic energy, as if the spins were coupled to an underlying lattice. 

\textcolor{black}{The work is organized as follows. In Section \ref{thermostat}, we introduce the new conservative thermostat. In Section \ref{numerics}, we test the new method on the Heisenberg antiferromagnet and obtain the correct critical exponents, spin wave dispersion law and spin wave softening. Finally, we show in Section \ref{connection} that the relaxation coefficient of the stochastic thermostat can be  qualitatively and quantitatively connected to the microscopic parameters of the spin-lattice coupling.}

\textcolor{black}{
\section{Discrete Laplacian Thermostat}\label{thermostat}
In this Section, partly taking inspiration from the dynamical mesoscopic equations of conserved fields, we will devise a way to write a conservative thermostat for a discrete spin dynamics. As we shall see, the key objects to achieve this result will be the discrete Laplacian and the incidence matrix; for a discussion of both these quantities in the context of graph theory see \cite{gross2018graph}.
}

\textcolor{black}{
\subsection{Microcanonical Spin Dynamics}
We consider a system of $N$ spins, $\sigma_i^\mu$, with $i=1,\dots, N$ and $\mu=1,\dots, d$, obeying the Poisson brackets,
\begin{equation} 
\lbrace \sigma_i ^\mu , \sigma_j^\nu \rbrace = \hbar^{-1}  \sum_{\rho=1}^d\varepsilon_{\mu \nu \rho}\, \sigma_i^\rho \, \delta_{ij} \ ,
\label{poisson}
\end{equation}
where $\varepsilon_{\mu \nu \rho}$ is the Levi-Civita antisymmetric tensor. Microcanonical SD amounts to integrate the reversible equations of motion,
\begin{equation}
\frac{d \boldsymbol{\sigma}_i}{dt} = \lbrace H, \boldsymbol{\sigma}_i \rbrace  \ ,
\label{SD}
\end{equation}
which naturally conserve the energy $H$.
We consider the case in which also the total magnetization, 
\begin{equation}
\boldsymbol{M}=\sum_{i=1}^N\boldsymbol{\sigma}_i \ ,
\end{equation}
is a constant of motion, 
\begin{equation}
\frac{d\boldsymbol{M}}{dt}=0 \ ,
\end{equation} 
which, of course, amounts to require,
\begin{equation}
\lbrace H, \boldsymbol{M}\rbrace=0 \ .
\end{equation} 
As we discussed in the Introduction, the only way to thermalize a microcanonical SD simulation is to start the numerical experiment from an initial condition previously thermalized at temperature $T$ with a canonical non-conservative dynamics, typically Montecarlo \cite{Landau2000}. We want to change that; our aim is to add to the microcanonical SD dynamics \eqref{SD} some new irreversible relaxational terms, i.e. a {\it stochastic thermostat}, which relaxes the energy $H$, while at the same time conserving the magnetization $\MM$. 
}

\textcolor{black}{
\subsection{Inspiration from dynamical field theory}
Within dynamical field theory there is a standard method to achieve the equilibration of a field $\varphi(x,t)$, subject to the constraint that its space integral is a constant of motion; this method consists in adding to the reversible parts of the dynamics, an irreversible relaxational force and a stochastic noise linked to each other by a kinetic coefficient proportional to the Laplacian \cite{HH}. More precisely, we can write, 
\begin{equation}
\frac{\partial \varphi}{\partial t} = \mathrm{reversible \ terms} - \Gamma \frac{\delta \cal H}{\delta\varphi} + \xi \ ,
\label{baccala}
\end{equation}
where the noise correlator satisfies the equilibrium condition, 
\begin{equation}
\langle \xi(x,t) \xi(x',t') \rangle = 2 \Gamma\ \delta(t-t')\delta(x-x') \ ,
\label{pippazza}
\end{equation}
and where (crucially) the kinetic coefficient $\Gamma$ is given by, 
\begin{equation}
\Gamma = -\lambda\nabla^2 \ .
\label{lallazione}
 \end{equation}
The positive parameter $\lambda$ is usually called transport coefficient; notice that also the kinetic coefficient $\Gamma$ is positive, as the continuous Laplacian is a negative-definite operator. To see why this method works, it is sufficient to go to Fourier space, where $ -\lambda\nabla^2 \to \lambda k^2$, so that the space integral of the field -- namely the mode $\varphi(k,t)$ at $k=0$ -- is automatically conserved, both by the relaxational force and by the noise.
}

\textcolor{black}{
Using the standard terminology of Hohenberg and Halperin \cite{HH}, this method is used in Model B (spinodal decomposition), in Models E and F (superfluid helium), in Model G (quantum antiferromagnet), and in Model J (isotropic quantum ferromagnet). Moreover, a conserved noise with the form of \eqref{pippazza}-\eqref{lallazione} is used in the context of non-equilibrium theories, in particular in the case of the conserved KPZ equation \cite{sun1989}. 
We want to take inspiration from this mesoscopic continuous case to devise a conservative relaxational dynamics that works also in the  microscopic discrete case. We stress however that we are {\it not} discretizing equations \eqref{baccala}-\eqref{lallazione} in any concrete way; we will simply try and export the physical mechanism of conservation used in the continuous case to the discrete setup. 
}

\textcolor{black}{
\subsection{The role of the discrete Laplacian}
\label{burino}
Transposing to the discrete case the idea behind the mesoscopic conservative dynamics \eqref{baccala}-\eqref{lallazione} does not seem too difficult, given that there exists a very well-known discrete version of the Laplacian operator; this matrix, that we shall call $\Lambda_{ij}$, is defined in the following way \cite{gross2018graph}, 
\begin{equation}
\Lambda_{ij} = -n_{ij}+\delta_{ij}\sum_k n_{ik} \ , 
\end{equation}
where $n_{ij}$ is the adjacency matrix defining the lattice's topological structure: $n_{ij}=1$ if two sites interact with each other, $n_{ij}=0$ otherwise; we will assume a symmetric interaction network, so that also the Laplacian is symmetric. Notice that -- at variance with its continuous counterpart -- the discrete Laplacian is a positive-definite matrix, i.e. $\Lambda \sim -\nabla^2$ (to make this correspondence dimensionally consistent we should include the square of the lattice spacing; however - as we have already remarked -- we are not pursuing an actual discretization of the continuous case, but just using it as a conceptual guideline).
One of the crucial features of the Laplacian is the fact that it has a {\it zero mode}, namely, 
\begin{equation}
\sum_i\Lambda_{ij}=0 \ .
\end{equation}
We can exploit this property and directly mimic equations \eqref{baccala}-\eqref{lallazione}, so to achieve a relaxational dynamics of the discrete spins, which at the same time enforces the conservation law. We propose to do this by writing the following canonical stochastic equations,
\begin{equation}
\frac{d \boldsymbol{\sigma}_i}{dt} = \lbrace H, \boldsymbol{\sigma}_i \rbrace -  
\hbar^{-1} \lambda \sum_j  \; \Lambda_{ij} \, \frac{\partial H}{\partial \boldsymbol{\sigma}_j} + \boldsymbol{\xi}_i  \  ,
\label{glip}
\end{equation}
where -- in order to achieve thermal equilibrium -- the dimensionless relaxation coefficient $\lambda$ and the noise $\xxi_i$ must satisfy the Fluctuation-Dissipation (FD) relation,
\begin{equation}
\langle{\xi}^\mu_i(t) {\xi}^\nu_j(t') \rangle = 2 k_\mathrm{B}T \; \hbar^{-1} \lambda \;\Lambda_{ij} \,\delta_{\mu\nu}\,\delta(t-t') \ .
\label{banzai}
\end{equation}
Let us show that dynamics \eqref{glip}-\eqref{banzai} conserves the total magnetization; we recall that $\lbrace H, \boldsymbol{M}\rbrace=0$ and that $\sum_i \Lambda_{ij}=0$; we therefore have, 
\begin{equation}
\frac{d\boldsymbol{M}}{dt} = \sum_i\frac{d \boldsymbol{\sigma}_i}{dt} = \sum_i \boldsymbol{\xi}_i \ ,
\end{equation}
but from \eqref{banzai} we see that the random variable $\sum_i \boldsymbol{\xi}_i$ has zero mean and zero variance, hence it must be identically zero for each realization of $\boldsymbol{\xi}_i$, finally proving that, 
\begin{equation}
\frac{d\boldsymbol{M}}{dt} =0 \ .
\end{equation}
On the other hand, the new irreversible relaxational term proportional to the `force', $\partial_{\boldsymbol{\sigma}} H$, pushes the spins to relax towards the configuration that minimizes the Hamiltonian, so that the energy converges to its equilibrium value at temperature $T$. Indeed, a standard Fokker-Planck argument \cite{Zwanzig} shows that the equilibrium probability distribution generated by \eqref{glip}-\eqref{banzai} is the Gibbs-Boltzmann canonical ensemble, 
\beq
P(\boldsymbol{\sigma}) = \exp(-H(\boldsymbol{\sigma})/k_\mathrm{B}T)/Z \ .
\eeq
We therefore have a new canonical stochastic dynamics, in which the microscopic spins are thermalized at temperature $T$ and the total magnetization is conserved.  And yet, we still have some more work to do.}

\begin{figure}[t]
\centering
\includegraphics[width=0.40\textwidth]{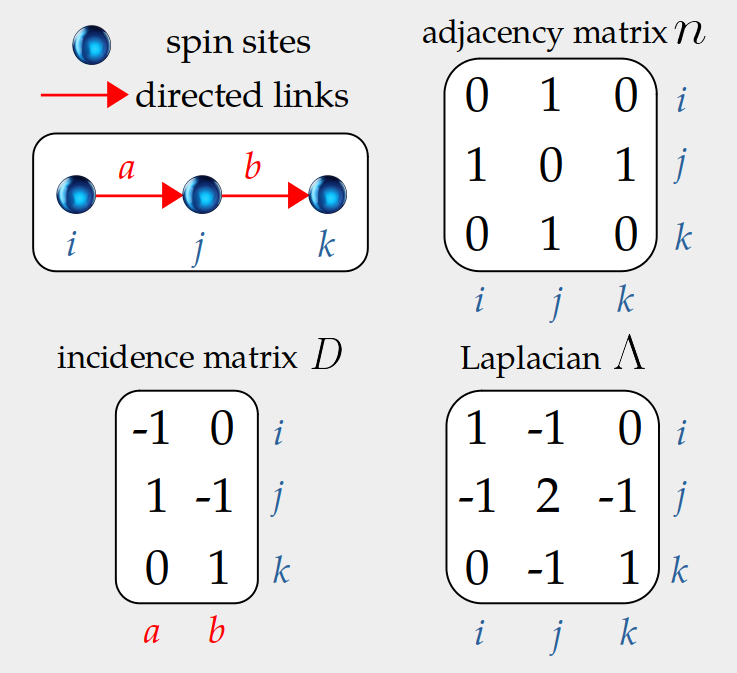}
\caption{Schematic view of the adjacency matrix $n$, incidence matrix $D$, and discrete Laplacian $\Lambda$ in a very simple lattice.}
\label{fig:scheme}
\end{figure}

\textcolor{black}{
\subsection{Incidence matrix noise}
\label{nana}
Up to now the method simply mimics the continuous case, but of course the real problem is how to produce a noise $\xxi_i$ whose correlator is proportional to the discrete Laplacian, $\Lambda_{ij}$, as required by equation \eqref{banzai}. Let us see how we can solve this problem.
\subsubsection{The complicated way}
If we insist working exclusively in the space of sites, the first possibility that comes to mind is to try and find the matrix $D_{ij}$ whose square (over the sites) is the Laplacian,
\beq 
\sum_{\mathrm{sites\ } k} D_{ik}D_{kj} = \Lambda_{ij}  \ ,
\label{kordoba}
\eeq
and then define the conservative noise as, $\xi_i=\sum_k D_{ik}\epsilon_k$, with $\langle \epsilon_k\epsilon_l\rangle\sim\delta_{kl}$, hence
giving $\langle\xi_i \xi_j\rangle \sim \Lambda_{ij}$, as desired. 
}

\textcolor{black}{
Although at first sight this seems feasible, it is in fact a dead end.
The problem with this method is that solving equation \eqref{kordoba} is far from straightforward: the matrix $D_{ij}$ heavily depends on the specific nature of the lattice, because one needs to go through the explicit form of the eigenvalues and eigenvectors of the discrete Laplacian to find it; in fact, one finds, 
\begin{equation}
D_{ij} = \sum_{\boldsymbol q} w_i^{\boldsymbol q} (w_j^{\boldsymbol q})^* \sqrt{\lambda_{\boldsymbol q}}
\label{mazinga}
\end{equation}
where $w_i^{\boldsymbol q}$ are the eigenvectors of the discrete Laplacian matrix $\Lambda$, with $\lambda_{\boldsymbol q}$ the corresponding eigenvalues.
This form of $D_{ij}$ is problematic, because the spectrum of the Laplacian is known analytically only for a limited number of regular lattices, and even in these cases only with periodic boundary conditions, while we would like to have a method that works irrespective of the specific topology of the lattice, and of its boundary conditions. Moreover, even in those cases where the Laplacian spectrum is exactly known and the matrix $D_{ij}$ can be calculated, its mathematical expression is extremely cumbersome, even for the simplest lattices. An even more serious problem is that of {\it locality}: in general, the matrix $D_{ij}$ in \eqref{mazinga} is non-zero even for non-interacting sites, namely for pairs of sites for which the adjacency matrix is zero, $n_{ij}=0$; this means that the conserved noise, $\xi_i=\sum_k D_{ik}\epsilon_k$, connects sites that were not directly interacting in the original Hamiltonian, which seems unnatural, to say the least. 
}

\textcolor{black}{
We want to develop a method that is local and that employs noting more complicated than the plain adjacency matrix itself, $n_{ij}$, and that possibly entails no calculations what-so-ever, neither hard, nor easy. Once again, field theory comes to our rescue.}

\textcolor{black}{
\subsubsection{The simple way}
In the continuous case, the fact that the noise variance is proportional to the Laplacian suggests that, in some way, the noise must be proportional to a gradient, $\xi\sim\nabla$. Fortunately, a simple discrete equivalent of the gradient does exist in graph theory: interestingly, it is a matrix defined in the space of {\it sites and links}, rather than of sites only. Let us label the sites of the lattice with $\lbrace i,j,\dots\rbrace$ and the links with $\lbrace a, b,\dots\rbrace$. The {\it incidence matrix}, $D_{ia}$, is constructed as follows  \cite{gross2018graph}: after arbitrarily assigning a direction to each link $a$, we set $D_{ia}=+1$ if $i$ is at the end of $a$,  $D_{ia}=-1$ if $i$ is at the origin of $a$, and $D_{ia}=0$ if site $i$ does not belong to $a$ (Fig.\ref{fig:scheme}); note that, by construction, we have, 
\beq
\sum_{\mathrm{sites\ } i} D_{ia} = 0 \ .
\label{piscotto}
\eeq
A little reflection immediately shows in what sense the incidence matrix is the discrete equivalent of the gradient: the `derivative' of a discrete set of variables $\{\sigma_i\}$ over link $a$ can now be written as,
\beq
[\nabla \sigma]_a = \sum_{\mathrm{sites\ } i} D_{ia} \sigma_i \ ,
\label{costarossa}
\eeq
in view of which, relation \eqref{piscotto} simply expresses the obvious, i.e. that the derivative of a constant is zero. Notice also that the arbitrariness in the definition of $D_{ia}$ due to the arbitrary choice of the directions of the links, reflects the inevitable arbitrariness in the definition of the derivative on a general discrete lattice. 
}

\textcolor{black}{
We can now state the crucial property of the incidence matrix, namely that its square {\it over the links} is equal to the discrete Laplacian \cite{gross2018graph}, 
\begin{equation}
\sum_{\mathrm{links\ } a} D_{ia}D^\mathrm{T}_{aj} = \Lambda_{ij}  \ .
\label{zimbra}
\end{equation}
What we have to do now seems clear: we need to switch from a noise defined on the {\it sites}, to a noise defined on the {\it links}. 
More precisely, on each link $a$ we define a standard $\delta$-correlated Gaussian noise, $\boldsymbol{\epsilon}_a$, with variance, 
\begin{equation}
\langle {\epsilon}^\mu_a(t) {\epsilon}^\nu_b(t') \rangle = 2k_\mathrm{B}T \; \hbar^{-1} \lambda\; \delta_{ab} \,\delta_{\mu\nu} \, \delta(t-t') \ ,
\label{pino}
\end{equation}
so that the site noise acting on each spin $i$ can finally be constructed as,
\begin{equation}
\boldsymbol{\xi}_i(t)= \sum_a D_{ia} \, \boldsymbol{\epsilon}_a(t) \ ,
\label{zumba}
\end{equation}
which gives discrete flesh to the idea that the conserved noise is proportional to a gradient, $\xi\sim\nabla$.
Let us compute the variance of this new noise, 
\begin{eqnarray}
 \langle{\xi}^\mu_i(t) {\xi}^\nu_j(t') \rangle &=&
\sum_{ab} D_{ia} D_{jb} \, \langle \epsilon^\mu_a(t) \epsilon^\nu_b(t') \rangle
\nonumber
\\
 &=& \sum_{a} D_{ia} D^\mathrm{T}_{aj} \, 2k_\mathrm{B}T \; \hbar^{-1} \lambda \,\delta_{\mu\nu} \, \delta(t-t') 
\nonumber
 \\
&=&2k_\mathrm{B}T \; \hbar^{-1} \lambda\,  \Lambda_{ij} \,\delta_{\mu\nu} \, \delta(t-t') \ ,
\end{eqnarray}
so that we recover exactly the desired expression, equation \eqref{banzai}. Moreover, we see from equation \eqref{zumba} that, within this construction, the site noise, $\xxi_i$, is the sum of all the link noises, $\boldsymbol{\epsilon}_a$, incident on that site; because by construction $\sum_i D_{ia} = 0$, from \eqref{zumba} we have that,
\beq 
\sum_i \boldsymbol{\xi}_i = 0 \ , 
\eeq
which makes it even more apparent the fact that the new noise conserves the total magnetization in \eqref{glip}.}

\textcolor{black}{
\subsection{Summary of the Discrete Laplacian Thermostat}
We have finally derived a closed set of equations specifying the canonical stochastic dynamics of a spin system with conserved magnetization. Because of the rather lengthy derivation, we summarize here the new canonical equations,
\begin{eqnarray}
\frac{d \boldsymbol{\sigma}_i}{dt} = \lbrace H, \boldsymbol{\sigma}_i \rbrace  && \, - \,   
\hbar^{-1} \lambda \sum_j  \; \Lambda_{ij} \, \frac{\partial H}{\partial \boldsymbol{\sigma}_j}  + \boldsymbol{\xi}_i  \  ,
\label{DLT1}
\\
\boldsymbol{\xi}_i(t)\, &&= \sum_a D_{ia} \, \boldsymbol{\epsilon}_a(t) \ ,
\label{DLT2}
\\
\langle {\epsilon}^\mu_a(t) {\epsilon}^\nu_b(t') \rangle \,  &&= 2k_\mathrm{B}T \; \hbar^{-1} \lambda\; \delta_{ab} \,\delta_{\mu\nu} \, \delta(t-t') \ ,
\label{DLT3}
\end{eqnarray}
where $\Lambda_{ij}$ is the discrete Laplacian and $D_{ia}$ is the incidence matrix. We call this method Discrete Laplacian Thermostat (DLT).
}

It is important to stress that for $\lambda=0$ the DLT canonical dynamics \eqref{DLT1}-\eqref{DLT3} becomes identical to microcanonical SD \eqref{SD}, where the energy does not relax; hence, we expect the relaxation coefficient $\lambda$ to be related to the inverse of the time-scale of energy thermalization. Apart from this, we will show that the relaxation coefficient does not impact on the qualitative features of the system: both the static and dynamic universality classes are unchanged, and the classic spin-wave phenomenology is correctly reproduced by DLT.

\textcolor{black}{
We conclude this Section with a notational clarification. After equations \eqref{baccala}, \eqref{pippazza} and \eqref{lallazione}, 
one could have expected us to call $\lambda$ the `transport coefficient', as in the field theory context \cite{HH}.
However, the terminology `transport coefficient', as well as `kinetic coefficient', belongs to the very specific context of hydrodynamics, which is a {\it coarse-grained} theory. Here, on the other hand, we are dealing with microscopic dynamic equations. Moreover, under coarse-graining, the microscopic spins may in general give rise to both conserved and non-conserved hydrodynamic fields, depending on the specific model \cite{HH, tauber2014critical}, so that 
the {\it microscopic} parameter $\lambda$ could contribute at the {\it coarse-grained} level both to the transport coefficient of a conserved field and to the kinetic coefficient of a non-conserved field. Therefore, we prefer to adopt a more neutral name for the microscopic parameter $\lambda$, and call it `relaxation' coefficient.
}

\begin{figure*}[t]
\centering
\includegraphics[width=0.99\textwidth]{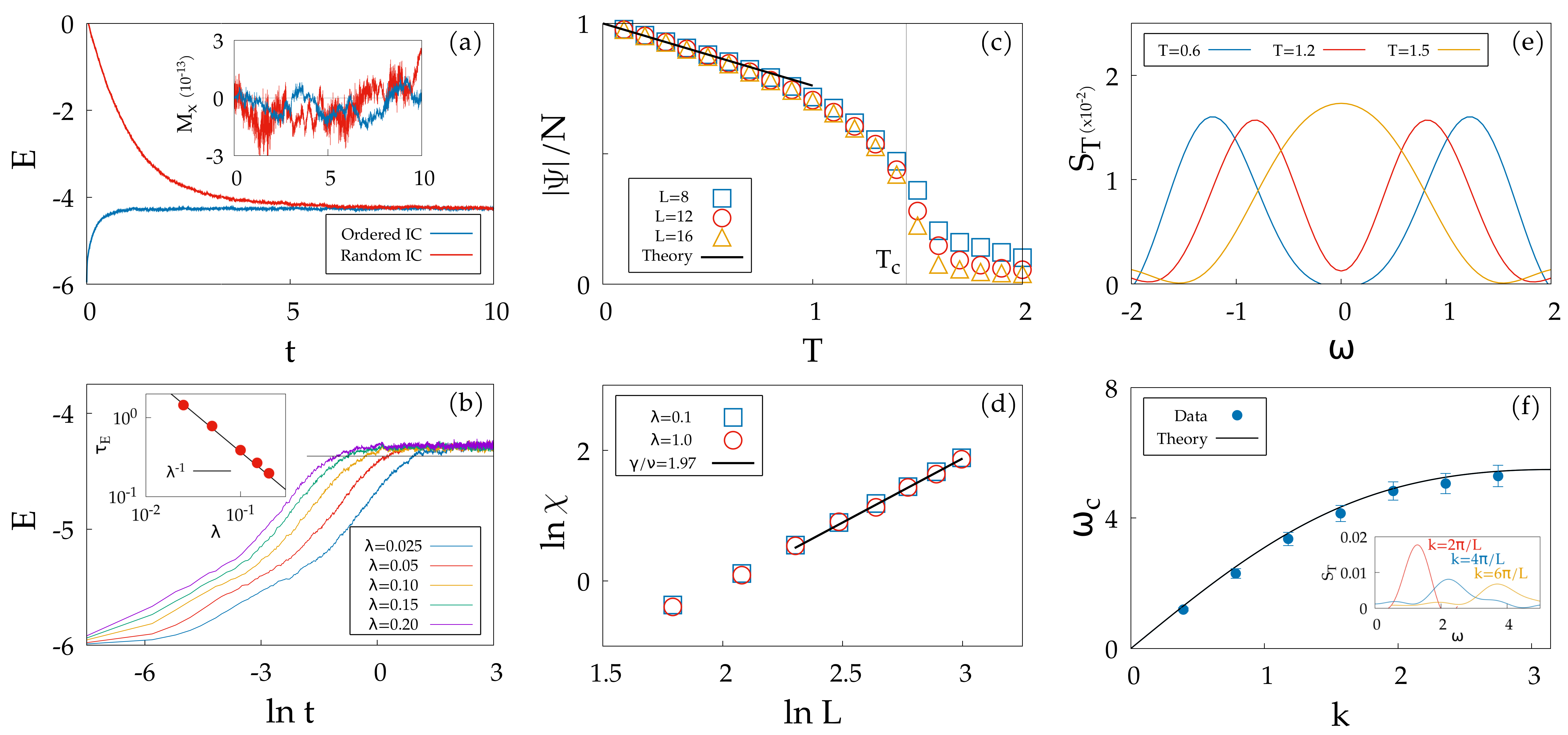}
\caption{
{\bf a.} Time evolution of the energy per spin, for  $L=16$, $T=0.6$ and $\lambda=0.1$, for ordered and random initial conditions. Inset:  $x$ component of the magnetization, fluctuating around the 13th decimal digit.
{\bf b.}  Energy relaxation (ordered initial conditions) at various values of $\lambda$. Inset: the energy thermalization time $\tau_E$ -- defined by the crossing to an arbitrary value of $E$ close to its asymptotic limit -- goes like, $\tau_E\sim\lambda^{-1}$. 
{\bf c.}  Modulus of the equilibrium staggered magnetization per spin vs temperature, at different sizes. The line corresponds to the low-$T$ linear spin-waves regime. 
{\bf d.}  Static susceptibility vs size, in the scaling regime, for two values of the relaxation coefficient $\lambda$; the line is the fit to the exact critical exponent.
{\bf e.} Tranverse scattering function $S_T$ vs frequency, for $k=\frac{2 \pi}{L}$, with $L=16$ and $\lambda=0.1$, at different temperatures. 
{\bf f.}  Dispersion law: characteristic spin-wave frequency vs wavevector for $T=0.6$ with $L=16$ and $\lambda=0.1$; the line is not a fit, but the actual analytic result (\ref{eq:dis}). Inset: spin wave peak at various wavevectors $k$.
}
\label{fig:2_3}
\end{figure*}

\section{Testing DLT in the Heisenberg Antiferromagnet} 
\label{numerics}

We numerically test DLT on the classical Heisenberg antiferromagnet. The Hamiltonian is,
\begin{equation}
     H= \frac{J}{2} \sum_{kl}  n_{kl}  \ \boldsymbol{\sigma}_k \cdot \boldsymbol{\sigma}_l  \ ,
     \label{stankonia}
\end{equation}
where $J>0$ and $n_{kl}$ corresponds to a $d=3$ cubic lattice of side $L$ with PBC. The Hamiltonian is rotationally invariant in the absence of an external field, so that $\lbrace H,\MM \rbrace =0$ and the global magnetization is conserved, $\dot\MM=0$. The order parameter is the non-conserved staggered magnetization, $\boldsymbol{\Psi}=\sum_i {\pi_i} \;\boldsymbol{\sigma}_i$, where $\pi_i=\pm 1$ is the parity of site $i$. 

By plugging Hamiltonian \eqref{stankonia} into the DLT equation \eqref{DLT1}, and after using Poisson's relations \eqref{poisson}, we obtain,
\begin{equation}
     \frac{d \boldsymbol{\sigma}_i}{dt} =
     \hbar^{-1} \frac{\partial H}{\partial \boldsymbol{\sigma_i}} \times \boldsymbol{\sigma}_i 
     - \hbar^{-1} \lambda \, \sum_{j}  \Lambda_{ij} \frac{\partial H}{\partial \boldsymbol{\sigma_j}}
     +  \boldsymbol{\xi}_i \ \ ,
     \label{eq:relax_a}
\end{equation}
where $\boldsymbol{\xi}_i$ is given by \eqref{DLT2} and \eqref{DLT3}. In order to fix the norm of the spins one could use a Lagrange multiplier, which would however slow down the simulation; instead, we use a single-particle  potential suppressing norm fluctuations (see Appendix A1 for more details). We have performed simulations of system sizes from $L^3=216$ up to $L^3=8000$ (the lattice spacing is $\ell=1$).  We work in units such that $k_\mathrm{B}=1$ and $\hbar=1$; we also choose units such that $J=1$, hence we are effectively measuring time in units of $J^{-1}$. 

\textcolor{black}{
\subsection{Numerical integrator}
We need to make an important technical remark here. Standard, microcanonical SD employs a non-stochastic fourth-order Runge-Kutta integrator with time-step $\Delta t=5 \cdot 10^{-4}$; of course, in any reversible microcanonical dynamics, in which energy must be conserved, it is important that the integrator is highly accurate, lest the conservation of energy is violated, which does not bode well for a microcanonical dynamics. But if we add a stochastic thermostat to the dynamics, the energy is no longer conserved, so that the tiny inaccuracies in the integration of the reversible part, which would cause a violation of energy conservation in the microcanonical case, become now irrelevant compared to the relatively huge fluctuations of the energy caused by the irreversible stochastic term; hence, what would normally happen when we switch from a non-stochastic microcanonical dynamics to a stochastic canonical one, is that we should also switch from a non-stochastic highly accurate integrator to a stochastic one.
}

\textcolor{black}{
But in our case, we want to be able to precisely recover the microcanonical SD dynamics in the limit $\lambda\to0$; while the case at exactly $\lambda=0$ could be studied by switching back to a non-stochastic integrator, this is not possible for small values of $\lambda$, when inaccuracies in the integration of the reversible term are {\it not} irrelevant compared to the energy fluctuations caused by the irreversible stochastic term. Therefore, the deterministic term must still be integrated with high accuracy, to correctly reproduce also the case of small relaxation coefficient. This is the reason why, even though we are dealing with a stochastic differential equation, we employ the same non-stochastic fourth-order Runge-Kutta integrator as in standard SD. 
}

\textcolor{black}{
On the other hand, by construction, both the reversible term and the irreversible thermostat satisfy {\it exactly} the constraint $\dot\MM=0$ (see Sections \ref{burino} and \ref{nana}), independently from the accuracy of the integrator, simply thanks to the antisymmetric form of the equations; hence, the conservation law of the magnetization is immune from all this.
}

\subsection{Conservation, transition, susceptibility}
\label{ponyo}
In Fig.\ref{fig:2_3}a we show that DLT conserves the magnetization with very high precision, while relaxing the energy, both when starting from random initial conditions, $\ppsi\sim0$, $\MM\sim0$, and when starting from ordered initial conditions, $|\ppsi|=N$, $\MM=0$;
to avoid slowing down due to coarsening we used ordered initial conditions in the rest of our study.
Fig.\ref{fig:2_3}b shows that energy thermalization is quicker the larger is the relaxation coefficient, $\lambda$; the energy is not a critical variable, hence its thermalization time, $\tau_E$, is a harmless microscopic scale of the system; we find $\tau_E \sim \lambda^{-1}$, hence confirming the expectation that the relaxation coefficient is the inverse of the energy time scale (had we not set $J=1$, we would have $\tau_E \sim (J\lambda)^{-1}$).

The antiferromagnet has a continuous phase transition at $T_c=1.446$ \cite{Tc_2001,Landau1993_1}, which DLT captures well (Fig.\ref{fig:2_3}c). Notice that the low-$T$ linear behaviour of the modulus of the staggered magnetization, $1- \Psi/N \sim T$, predicted by the theory \cite{hydrosw}, is also correctly reproduced by DLT.  

We test the static critical behaviour by studying the susceptibility, which satisfies the finite-size scaling relation, $\chi = \xi^{\gamma/\nu}g(L/\xi)$, where $g(x)$ is a scaling function; we probe the scale-free regime by selecting at each size $L$ the temperature $T_c(L)$ at which $\chi$ is maximal (see Appendix A2 for a detailed description of the procedure), hence we obtain $\xi \sim L$, and therefore $\chi \sim L^{\gamma/\nu}$; the fit to the theoretical exponent \cite{gammanu} is quite satisfactory (Fig.\ref{fig:2_3}d).

\subsection{Spin waves and their dispersion relation}
The low temperature regime of antiferromagnets  is dominated by spin waves \cite{Dyson}, which are the quintessential consequence of the symmetry and conservation law; hence it is important to check how DLT performs in relation to them. The transverse  scattering function $S_T(k,\omega)$ of the staggered magnetization -- which is computed following \cite{Landau1996} -- is reported in Fig.\ref{fig:2_3}e: as expected, above the critical temperature there is a simple diffusive peak, while two spin-wave peaks at $\pm \omega_c$ emerge for $T < T_c$. 

\begin{figure}[t]
\centering
\includegraphics[width=0.5\textwidth]{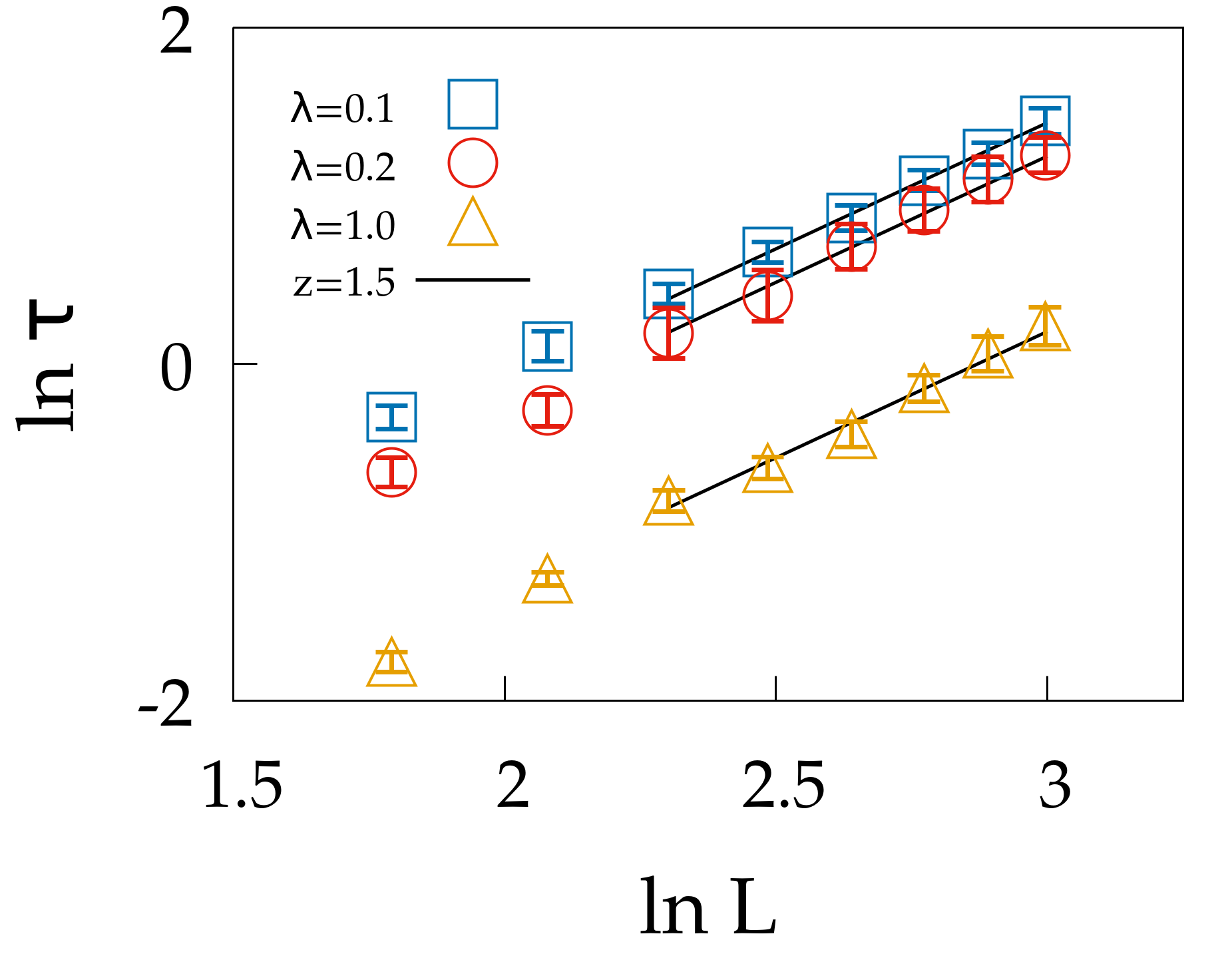}
\caption{Relaxation time $\tau$ as a function of the system's size $L$, for different values of $\lambda$. The lines indicates fits to the exact critical exponent, $z=1.5$, over the five largest sizes.
}
\label{fig:4}
\end{figure}
In the spin wave phase, the characteristic frequency $\omega_c$ depends on the wavevector $k$ according to the exact dispersion relation (see Appendix B for the full derivation), 
\begin{equation}
\omega_{c}(k)= 4J \sqrt{d}\; \sin(k\ell/2) \sqrt{1- (1/d) \sin(k\ell/2)^2} \ ,
\label{eq:dis}
\end{equation}
which is fairly well reproduced by DLT, considering that no fitting parameters what-so-ever are used (see Fig.\ref{fig:2_3}f).

\subsection{Critical dynamics and the exponent $z$}
As fundamental as the existence of spin waves is the emergence of critical slowing down at the phase transition: 
in the bulk, the relaxation time of the order parameter at $k=0$ diverges at the critical point as a power law of the correlation length, $\tau \sim \xi^z$ \cite{HH}. An exact renormalization group calculation of the dynamic critical exponent  yields $z = 1.5$ for the Heisenberg antiferromagnet in $d=3$ \cite{mazenko_rg}, a result that has been confirmed both by numerical simulations \cite{Landau2003} and by experiments on perovskites \cite{tucciarone,zexp}.

\textcolor{black}{Notice that $z = 1.5$ differs very significantly from the value $z\approx 2$ of the universality class of standard non-conserved ferromagnets (as the Ising model), also called Model A class in the classification of Hohenberg-Halperin \cite{HH}}. The profound reason for this difference is the connection between symmetry and conservation law: if the critical dynamics of the antiferromagnet is studied through a Montecarlo method, one obtains $z\approx 2$; interestingly, it is not only standard non-conserved Montecarlo that fails to yield the correct dynamical critical exponent \cite{z2}, but also Kawasaki Montecarlo (KM) fails:  although KM conserves $\MM$, this conservation has no effects on the dynamics of $\ppsi$, hence giving $z\approx 2$; as we have already noted, it is not the conservation {\it per se}, but the deep dynamical connection between symmetry and conservation law that yields the correct universality class.

\begin{figure}[t]
\centering
\includegraphics[width=0.5\textwidth]{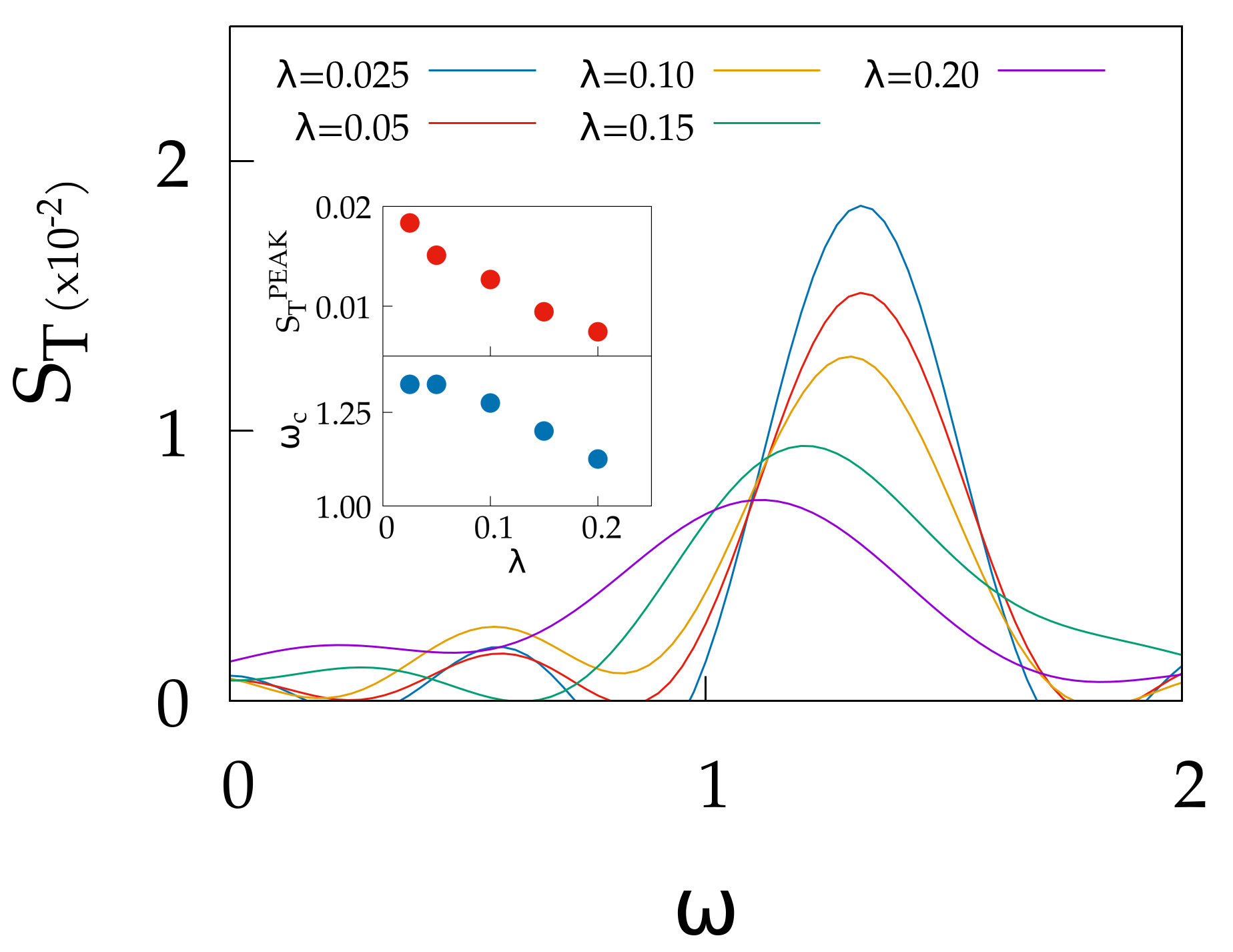}
\caption{The intensity and the position of the spin wave peak in the low temperature phase decrease when the relaxation coefficient $\lambda$ increases ($T=0.3$ and $L=16$). Inset: evolution of the peak's height and frequency with $\lambda$.
}
\label{fig:5}
\end{figure}
\textcolor{black}{
To calculate the critical exponent $z$ we use dynamical scaling \cite{HH_scaling}, according to which the relaxation time has the form,
\begin{equation}
    \tau_k=k^{-z}g(k\xi) \ ,
\label{bubu}
\end{equation}
where $g(x)$ is a scaling function and all the dependence on the temperature $T$ goes into $\xi(T)$ (the precise definition of relaxation time is reported in Appendix A3). We now define the largest relaxation time, $\tau$, as that corresponding to the lowest mode, $k=\frac{2 \pi}{L}$; from \eqref{bubu} we obtain,
\begin{equation}
    \tau = L^z g(\xi/L) \ .
\label{baba}
\end{equation}
If for each size $L$ we work in the scale-free regime, namely at the pseudo-critical temperature, $T_c(L)$, where $\chi$ is maximal (see Section \ref{ponyo} and Appendix A2), we have $\xi(T_c(L)) \sim L$, hence, 
\begin{equation}
    \tau \sim L^z \ ,
\label{bibi}
\end{equation}
which is the relation we test, using sizes ranging from $L^3=1000$ to $L^3=8000$. The result of DLT is quite satisfactory (Fig.\ref{fig:4}): we find $z=1.47 \pm 0.07$ for $\lambda=0.1$, $z=1.56 \pm 0.08$ for $\lambda=0.2$, and $z=1.50 \pm 0.10$ for $\lambda=1.0$, all values consistent with the exact result, $z=1.5$. 
The tests of thermalization that we used to check that all simulations are actually at equilibrium are reported in Appendix A4.
}

Hence, the critical exponent $z$ does not depend on the relaxation coefficient $\lambda$, which therefore does not impact on the dynamic universality class of the model; the role of the relaxation coefficient is simply to shift the prefactor of $\tau$, which is reasonable, given the identification of $\lambda^{-1}$ as a non-critical time scale of the system.

\subsection{Spin wave softening and damping}
A feature over which the relaxation coefficient {\it does} have a nontrivial impact is the blunting of spin waves (Fig.\ref{fig:5}):
the larger is $\lambda$, the weaker is the spin wave peak (SW {\it damping} \cite{Sanger1994}) and the lower its characteristic frequency (SW {\it  softening} \cite{deymier2014phonon}).

 Damping and softening are typically caused by the coupling of the magnetic degrees of freedom (magnons) with the lattice vibrational degrees of freedom (phonons) \cite{woods2001magnon}, an interpretation confirmed by numerical simulations \cite{soft_I, soft_II}.
Hence, the fact that $\lambda$ is connected to softening suggests to try and link the relaxation coefficient of DLT to the microscopic details of the magnon-phonon system. This is what we do in the next Section.


\section{Connection between DLT and spin-lattice dynamics}
\label{connection}

\textcolor{black}{
The canonical DLT method has just one extra parameter compared to microcanonical SD, namely the relaxation coefficient $\lambda$, which -- as we have seen -- is essentially the inverse of the energy relaxation time; this makes sense, as in SD the energy does not relax, of course, and the limit $\lambda\to 0$ reproduces exactly the SD case. We also have found from the numerical simulations that $\lambda$ has an impact on the softening of spin waves, while no spin-wave softening can be observed in microcanonical SD. In actual physical systems, that is in systems where the spins are coupled to an  underlying atomic lattice, both these phenomena (energy relaxation and spin-wave softening) are obviously caused by the very interaction between spins and lattice. Therefore, it seems natural to seek a connection between the relaxation coefficient $\lambda$ of DLT and the physical parameters of the actual spin-lattice coupling. To do that, we need a model of the interaction between spins and atomic lattice.
}

\subsection{The simplest SD+MD dynamics}
 In the very simplest case the coupling between the spins and the underlying lattice can be described by the Hamiltonian \cite{Sanger1995,soft_IV}, 
\begin{eqnarray}\label{H_tot_main}
     H_\mathrm{TOT} &=& \frac{1}{2} \sum_{kl}J(r_{kl}) \; \boldsymbol{\sigma}_k \cdot \boldsymbol{\sigma}_l 
     \nonumber
     \\
    &+& \frac{1}{2} K  \sum_{kl} n_{kl} (\boldsymbol{u}_k-\boldsymbol{u}_l)^2 + \sum_k \frac{\boldsymbol{p}^2_k}{2m} \ ,
     \label{zulu}
\end{eqnarray}
where $r_{kl}=|\boldsymbol{r}_k - \boldsymbol{r}_l|$, while,
\beq
\boldsymbol{u}_k=( \boldsymbol{r}_k - \boldsymbol{r}_k^0) \ ,
\eeq
is the deviation of nucleus $k$ from its equilibrium position. 
The first term in $H_\mathrm{TOT}$ is the spin exchange interaction, while the second and third terms describe the vibrational excitations of the lattice. At variance with microcanonical SD, the spins' positions are not fixed and the spin exchange interaction, $J(r)$, depends on the distance. Within a microcanonical simulation of \eqref{zulu} (the SD+MD method described in the Introduction \cite{Phonon_I, Phonon_II}), the total magnetization $\MM$ is conserved, as well as the {\it total} energy $H_\mathrm{TOT}$ of the system; however, there is an energy exchange between phonons and magnons, so that the spin magnetic energy is not conserved and it relaxes to its thermal value. Hence, we can interpret phonons as a conservative thermostat coupled to magnons, which is exactly the role of DLT. 

\textcolor{black}{
To explore this analogy further, following  \cite{soft_IV,Sanger1994,Xiong2017} we expand the spin exchange coupling to first order, 
\begin{equation}
     J(r_{kl})= J_0 \; n_{kl}   -\alpha \; n_{kl} \; (\boldsymbol{u}_k-\boldsymbol{u}_l) \cdot \hat{\boldsymbol{e}}_{kl}   \  , 
     \label{fart}
\end{equation}
where,
\beq
J_0 = J(r_{kl}^0) \ ,
\eeq
becomes the spin-spin coupling constant at the leading order, while, 
\beq
\alpha = - J'(r_{kl}^0) \ , 
     \label{alphader}
\eeq
plays the role of the spin-lattice coupling constant; notice that $\alpha$ is positive, because the exchange interaction, $J(r)$, decreases with distance. Finally, in \eqref{fart} we have defined, 
\beq
\hat{\boldsymbol{e}}_{kl}=  (\boldsymbol{r}^0_{k}-\boldsymbol{r}^0_{l})/|\boldsymbol{r}^0_{k}-\boldsymbol{r}^0_{l}| \ .
\eeq
We can now write the equations of motion for all the degrees of freedom,
\begin{eqnarray}
     \frac{d \boldsymbol{\sigma}_i}{dt} &=& \hbar^{-1} J_0 \sum_k n_{ik}\ \boldsymbol{\sigma}_k \times \boldsymbol{\sigma}_i 
   \nonumber
     \\
     &-& \hbar^{-1} \alpha \sum_k n_{ik}\  \boldsymbol{\sigma}_k \times \boldsymbol{\sigma}_i  \ (\boldsymbol{u}_i-\boldsymbol{u}_k) \cdot \hat{\boldsymbol{e}}_{ik} \ ,
      \label{suka}
      \\
     \frac{d \boldsymbol{u}_i}{dt} &=& \frac{ \boldsymbol{p}_i}{m} \ , 
     \\
     \frac{d \boldsymbol{p}_i}{dt} &=& -2K \sum_k n_{ik}\ ( \boldsymbol{u}_i -\boldsymbol{u}_k)\
     \nonumber
     \\
     &&\quad\quad\quad\quad\quad +\, \alpha \sum_k n_{ik}\  (\boldsymbol{\sigma}_i \cdot \boldsymbol{\sigma}_k )  \  \hat{\boldsymbol{e}}_{ik} \ .
     \label{suka_fononi}
\end{eqnarray}
These equations conserve the total magnetisation thanks to the fact that the unit vector $\hat{\boldsymbol{e}}_{ik}$ is antisymmetric.
Compared to purely microcanonical SD, however, the spin equation, \eqref{suka}, has an extra spin-phonon term proportional to $\alpha$, which is responsible for relaxing the magnetic energy.
}

\textcolor{black}{
\subsection{Marginalizing the lattice degrees of freedom}
In order to find an effective canonical dynamics for the sole spin variables, we follow the rather standard path \cite{Zwanzig} of marginalizing the spin-phonon term 
over the dynamics of the phonon variables, $\lbrace\boldsymbol{u}_i, \boldsymbol{p}_i\rbrace$, using thermalized initial conditions. To do this, one must first rewrite the dynamical equations using the eigenvectors of the discrete Laplacian, which is somewhat lengthy and cumbersome; for this reason, the full details of the calculation are reported in Appendix C.
}

Once this marginalization is performed, a {\it relaxational} term and a {\it noise} term emerge, linked to each other by the FD relation and both conserving the total magnetization, thus acting as an effective thermostat very similar to DLT.  \textcolor{black}{We can write the formal solution of equation \eqref{suka_fononi} treating the spin term as an external driving force; then, the solution of \eqref{suka_fononi} can be plugged into \eqref{suka}, and after some manipulations and reasonable approximations  (described in Appendix C), we end up with, 
\begin{equation}
    \dfrac{d \boldsymbol{\sigma}_i}{dt} = \hbar^{-1} \dfrac{\partial H_{0}}{\partial \boldsymbol{\sigma}_i} \times \boldsymbol{\sigma}_i - \boldsymbol{\Xi}_i\left[\{\boldsymbol{\sigma}\}\right] +\boldsymbol{\xi}_i  \ .
    \label{ammappalo}
\end{equation}
The first term at the r.h.s. is the usual reversible force coming from Poisson's brackets structure, where,
\beq
H_0=\frac{J_0}{2}\sum_{kl}n_{kl}\, \boldsymbol{\sigma}_k\cdot\boldsymbol{\sigma}_l \ ,
\eeq
is the standard Heisenberg Hamiltonian, as in \eqref{stankonia}.
The second term in \eqref{ammappalo} is a relaxational force, which takes the form (see Appendix C),
\begin{equation}
\label{relaxational_kernel_micro_indices_main}
  \Xi_i^{\mu}=\hbar^{-1} \left(\dfrac{\alpha^2}{\hbar K}\right)\int_0^t dt' \sum_{j,\nu}  2   {\cal R}^{\mu\nu}_{ij}(t,t') \left( \dfrac{\partial H_{0}}{\partial \sigma^{\nu}_j} \right)_{t'} ,
\end{equation}
where ${\cal R}^{\mu\nu}_{ij}(t-t')$ is a dimensionless memory kernel, whose explicit form is specified in Appendix C (see equation \eqref{dimensionless_Kernel_micro}) and that has the following crucial property,
\beq
\sum_i {\cal R}^{\mu\nu}_{ij}(t-t') = 0 \ ,
\label{calippo}
\eeq
which guarantees the conservation of the total magnetization in \eqref{ammappalo}. The third term, $\boldsymbol{\xi}_i$, is a noise, with variance,
\begin{equation}
\langle {\xi}^\mu_i({t})  {\xi}^\nu_j({t'}) \rangle =
2k_\mathrm{B}T \; \hbar^{-1} \left(\frac{\alpha^2}{\hbar K}\right)  \; {\cal R}^{\mu\nu}_{ij}(t-t') \ .
\label{sandapa}
\end{equation}
Using the same line of thought of Section \ref{burino}, it is clear that this noise too conserves the total magnetization, because, thanks to \eqref{calippo} and \eqref{sandapa}, we have that $\sum_i \boldsymbol{\xi}_i = 0$ for each realization of the noise. All brackets, $\langle\cdot\rangle$, in the relations above indicate an average over the initial conditions for phonons, as they are random variables drawn from the Gibbs-Boltzmann distribution. We also observe that the FD relation is satisfied, since in both the relaxational and the noise term the same kernel ${\cal R}^{\mu\nu}_{ij}(t-t')$ appears.}

\textcolor{black}{
\subsection{A simple dimensional argument }
The derivation of equations \eqref{ammappalo}-\eqref{sandapa} -- derivation that we provide in great detail in Appendix C -- is lengthy and cumbersome. Hence, let us give to the reader a very simple dimensional argument to calculate the effect of the spin-lattice coupling on the dynamics of the spins, and show how DLT quite naturally emerges in this context.
}

\textcolor{black}{
If we go back to equation \eqref{suka}, we see that the first term at the r.h.s. is the same as in pure SD dynamics, therefore it will not be affected by the marginalization over the phonon variables, while the second term, proportional to the spin-lattice coupling constant $\alpha$, is the one that -- containing the phonon degrees of freedom -- will give rise both to the relaxational force, $\Xi$, and to the noise, $\xi$; let us concentrate on the latter, even though, of course, dimensionally the two are the same.
By dimensional analysis of the spin-phonon term in \eqref{suka} we get, 
\beq
\langle\xi^2 \rangle 
\sim \hbar^{-2} \alpha^2 \langle (\boldsymbol{u}_i-\boldsymbol{u}_k)^2\rangle \ ,
\label{toka}
\eeq
where -- again --  brackets, $\langle\cdot\rangle$, indicate an average over thermal phonons. By using equipartition of the phonons potential energy, we can rewrite \eqref{toka} as, 
\beq
\langle\xi^2 \rangle \sim k_\mathrm{B}T \; \hbar^{-1} \left(\frac{\alpha^2}{\hbar K}\right)  \ .
\label{tingle}
\eeq
We immediately see that, dimensionally, this relation is the same as \eqref{sandapa}; moreover, the only reasonable way to
generalize \eqref{tingle} to different sites, $i,j$, different coordinates, $\mu,\nu$, and different times, $t,t'$, without changing the dimensions, is indeed to introduce a dimensionless kernel ${\cal R}^{\mu\nu}_{ij}(t-t')$, which gives exactly equation \eqref{sandapa}; moreover, even if we did not do the explicit calculation of this kernel, we would still know that the conservation of magnetization that holds in the original equations \eqref{suka} must survive marginalization over the phonons, hence we must have that $\sum_i  {\cal R}^{\mu\nu}_{ij}=0$. 
}

\textcolor{black}{
Now that we have dimensionally justified the form of the effective noise variance \eqref{sandapa}, we can compare it with the DLT variance, that we
report here for clarity,
\begin{equation}
\langle{\xi}^\mu_i(t) {\xi}^\nu_j(t') \rangle = 2 k_\mathrm{B}T \; \hbar^{-1} \lambda \;\Lambda_{ij} \,\delta_{\mu\nu}\,\delta(t-t') \ .
\label{banzai2}
\end{equation}
Such comparison may suggest that we are almost done: we have obtained a non-Markovian non-isotropic version of the DLT noise and one may be tempted to identify the DLT relaxation coefficient $\lambda$ with $(\alpha^2/\hbar K)$; but that would be a dimensional mistake, because the kernel ${\cal R}^{\mu\nu}_{ij}(t-t')$ is dimensionless, while the combination $\Lambda_{ij} \,\delta_{\mu\nu}\,\delta(t-t')$ has the dimensions of the inverse of a time. 
To make progress, we introduce the memory time scale of the non-Markovian kernel,
\beq
\tau_\mathrm{m} \sim \int_{-\infty}^{\infty} ds \  {\cal R}^{\mu\nu}_{ij}(s) \ ,
\eeq
where we can forget about its possible $ij$ and $\mu\nu$ dependence as long as we are only proceeding through dimensional analysis. The advantage of having singled out $\tau_\mathrm{m}$ is that we can rewrite \eqref{sandapa} as,
\begin{equation}
\langle {\xi}^\mu_i({t})  {\xi}^\nu_j({t'}) \rangle =
2k_\mathrm{B}T \; \hbar^{-1} \left(\frac{\alpha^2}{\hbar K}\right) \tau_\mathrm{m} \ \left(\frac{ {\cal R}^{\mu\nu}_{ij}(t-t')}{\tau_\mathrm{m}}\right)   \ ,
\label{sandapaperone}
\end{equation}
where now the operator $({\cal R}^{\mu\nu}_{ij}/\tau_\mathrm{m})$ has the right dimensions to play the role of a `fatter' $\delta$-function, so that a direct comparison between \eqref{sandapaperone} and the DLT original noise \eqref{banzai2} finally gives,
\begin{equation}
\lambda =  \left(\frac{\alpha^2}{\hbar K}\right) \, \tau_\mathrm{m} \ , 
     \label{bonny}
\end{equation}
which is a quantitative relation between the relaxation coefficient of DLT and the microscopic parameters of the spin-lattice coupling.
}

\textcolor{black}{
These results can actually be obtained analitically, and we refer the reader to Appendix C for the details. There, we show that -- under some reasonable approximations -- the kernel is proportional to the discrete Laplacian,  
\beq
{\cal R}^{\mu\nu}_{ij}(t-t') \simeq \Lambda_{ij}\, \delta_{\mu\nu}\,  {\cal R}(t-t') \ , 
\label{semis}
\eeq
so that the effective noise after the phonons marginalization becomes indeed a non-Markovian version of the DLT noise, 
\begin{equation}
\langle {\xi}^\mu_i({t})  {\xi}^\nu_j({t'}) \rangle =
2k_\mathrm{B}T \; \hbar^{-1} \left(\frac{\alpha^2}{\hbar K}\right)   \Lambda_{ij}\, \delta_{\mu\nu}\,\; {\cal R}(t-t') \ .
\label{sandapapero}
\end{equation}
Then, the kosher way to proceed after this is to discretize the spin dynamics using a time interval $\Delta t \gg \tau_\mathrm{m}$, so that the dynamics becomes effectively Markovian over time scales comparable to the discretization scale; in this way we finally obtain a Markovian noise that can be directly compared with the time-discrete version of the DLT noise \eqref{banzai2}, thus giving again equation \eqref{bonny} (see -- as usual -- Appendix C for the details).
}

\textcolor{black}{
The relation we found between the DLT relaxation coefficient $\lambda$ and the parameters of the spin-lattice system -- equation \eqref{bonny} -- looks very sound at the qualitative level: when $K$ goes to infinity (i.e. for infinitely stiff -- namely fixed -- lattice) or when $\alpha$ goes to zero (no spin-lattice coupling), we obtain $\lambda\to 0$, correctly recovering microcanonical SD. Hence, the relaxation coefficient of DLT wraps into a single quantity all the parameters of the (possibly very complicated) interaction between spins and lattice: $\lambda$ is larger the stronger is the phonon-magnon coupling, and because this coupling is responsible for blunting spin waves \cite{soft_II}, we finally understand why softening and damping within DLT are the stronger the larger is $\lambda$. But can we trust relation \eqref{bonny} at the {\it quantitative} level?
}

\textcolor{black}{
First, let us discuss what happens when $\lambda$ is either too small or too large. As we have said, if there is no magnon-phonon interaction we obviously obtain $\lambda=0$; but we know that $1/\lambda$ is the energy relaxation time, which does not diverge in real antiferromagnets.
On the other hand, as we have seen from numerical simulations, too high a value of $\lambda$ would completely suppress spin waves, which are 
a quintessential feature of real antiferromagnets.
These considerations suggest that it may be possible to actually obtain a reasonable estimate of $\lambda$ by plugging into \eqref{bonny} the microscopic parameters of actual magnetic materials found in the literature \cite{Phonon_I, evans2014atomistic}.
We can estimate the strength of the phonon-magnon coupling, $\alpha$, from equation \eqref{alphader}, in particular by using the characteristic amplitude and width of standard forms of the function $J(r)$ (see -- for example -- Table A1 of \cite{Phonon_I}), getting $\alpha \sim 1.0 - 2.0\times10^{-11}$ J/m.
The lattice stiffness $K$ can be derived from sound velocities, atomic masses, and lattice spacings for magnetic systems, which gives $K \sim 40 - 50$ N/m. 
Estimating the value of the memory kernel time scale,  $\tau_m$, is more challenging, as it depends on the phonon dynamics -- itself coupled to the spins; it is known that the characteristic time scale of the phonon-phonon interaction is of the order of $1$ps, while the spin-phonon interaction has a characteristic scale of $100$ps (see \cite{Phonon_I}). It is, therefore, reasonable to expect that $\tau_m$ should be within this range.  
Hence, we obtain that the value of the (dimensionless) relaxation coefficient in real materials should lie in the range $ 0.02 < \lambda < 10 $, which is   consistent with the values of the relaxation coefficient used in our DLT simulation, where $\lambda$ was chosen between $0.1$ and $1$ (see Figs. \ref{fig:2_3}, \ref{fig:4} and \ref{fig:5}).}

\textcolor{black}{In conclusion, the connection between the effective canonical dynamics provided by DLT and the actual microcanonical dynamics of systems with spin-lattice coupling, seems to be deeper than a generic numerical shortcut: given a certain material, with a certain set of parameters describing its microcanonical SD+MD dynamics, it should be possible to {\it calculate} the DLT relaxation coefficient $\lambda$ through \eqref{bonny} and run an effective canonical dynamics appropriate for that specific material. 
}

\section{Conclusions}

We have introduced a new thermostat that is at the same time conservative and stochastic, thus preserving the true dynamics of the spins, yet making it canonical. The potential applications of DLT seem promising.
\textcolor{black}{ 
First of all, remaining at the equilibrium level, the method should be tested in other dynamical universality classes for which symmetries and conservation laws are important, as Model J (Heisenberg ferromagnet) and Models E/F (superfluid Helium) \cite{HH}. 
But DLT seems particularly promising for the study of out-of-equilibrium systems. As we anticipated in the Introduction, the method could be employed in the context of aging studies \cite{Tauber2019}, as quenches and dynamic changes of temperature become straightforward within the canonical dynamics. 
Moreover, DLT  -- or, more precisely, the conservative noise that we have introduce within DLT -- can be used for the study of inherently out-of-equilibrium systems, such as those that violate detailed balance; in these cases, Metropolis-like Montecarlo simulations -- whose dynamics is built upon the assumption of detailed balance -- cannot be employed. One outstanding example in this class of systems is the conserved KPZ equation \cite{sun1989}, which has non-thermal fluctuations. DLT noise could be an interesting new way to simulate such equations in real space.
}

\textcolor{black}{Of course, DLT is not free of limitations. As with any other thermostat, and despite the connection between the relaxation coefficient and the spin-lattice parameters, a substantial part of the physical information regarding the microscopic degrees of freedom is lost, resulting in an effective simplified canonical dynamics. Moreover, we have seen how DLT can deal with respecting {\it one} conservation law, that of the total order parameter; it remains to be seen whether, and how, DLT could be adapted or generalized to the case of several conservation laws simultaneously holding in the dynamics. 
}

\textcolor{black}{
An interesting issue for future studies is the possible emergence of dynamical crossovers related to the weak violation of the conservation law. 
In our -- much -- simplified microscopic model for spins and lattice we assumed that the only relevant interaction was the exchange interaction, which is isotropic, namely invariant under rotations in the internal space of spins, thus leading - through Noether's theorem - to the exact conservation law of the global magnetization. 
Within DLT, this exact conservation law is empowered by the fact that the relaxational force and the noise variance are proportional to the discrete Laplacian, $\Lambda_{ij}$.
However, anisotropic  interactions, such as spin-orbit or dipole-dipole interaction, can in some cases be relevant, thus violating the symmetry and the conservation law.  The interesting point is that we can easily generalize DLT to this cases, by adding a {\it non-conservative} component to the relaxation coefficient, namely,
\beq
\lambda\, \Lambda_{ij} \longrightarrow \lambda\, \Lambda_{ij} + \eta\, \delta_{ij} \ ,
\label{eta}
\eeq
leaving all the other relations untouched. Now the effective friction, $\eta$, dissipates the total magnetization, hence it can be used to describe those physical cases in which the original conservation law is hindered by all sorts of non-symmetric interactions.
When $\eta$ is very large compared to the conservative coefficient $\lambda$, the symmetry and conservation laws are completely washed away, and one recovers the physics of overdamped non-conservative systems \cite{gilbert2004}. But when anisotropic interactions are weak, the value of $\eta$ will be non-zero but small, so that the relative magnitude of the conservative relaxation coefficient, $\lambda$, vs the non-conservative one, $\eta$, may give rise to interesting finite-size dynamical crossovers. Such crossovers have been already studied in the field-theoretical context using the renormalization group \cite{cavagna2019renormalization}, so that the possibility to analyze at the microscopic level the interplay between conserved and non-conserved relaxation seems a promising direction of investigation in the general field of magnetism.
}

\textcolor{black}{
Finally, we notice that there might be a broader spectrum of applications of DLT beyond the realm of physical magnets. In biophysics there is growing experimental \cite{cons_exp_I,cons_exp_II}, numerical \cite{cons_spin} and theoretical \cite{cons_theo_I,NP2023} evidence that conservation laws are important in determining the collective dynamical properties of some natural systems. More precisely, it has been noted that, although the mechanisms of biological imitation between neighbours in a biological group can be described by borrowing the standard concepts of ferromagnetism  \cite{revbioIII}, an overdamped dynamics as in the standard non-conserved case (model A class \cite{HH}) does not reproduce some key experimental traits, as information propagation across the group and dynamic scaling \cite{cons_exp_I,cons_exp_II}. On the other hand, it has emerged that the dynamics of underdamped mode-coupling systems, that is systems where a conservation law is important (as the Model G class \cite{HH} of the antiferromagnet studied here), is more suited to the study of collective behaviour and it has a better agreement with biological experiments \cite{NP2023}; essentially, the behavioural {\it inertia} of the biological individuals forces the interaction rules to comply to a second order dynamics that can only emerge as a result of a conservation law \cite{cons_exp_I}. In the light of this, a conservative thermostat tailored on spin dynamics could become a relevant numerical tool also in physical biology. 
}

\section*{Acknowledgements}

This work was supported by ERC grant RG.BIO (n. 785932), and by grants PRIN-2020PFCXPE and FARE-INFO.BIO from MIUR. We thank  G. Pisegna and M. Scandolo for discussions, and F. Cecconi and J. Lorenzana for important comments upon reading the manuscript. We gratefully acknowledge compelling conversations with T.S. Grigera in the early stages of this work.

\bibliography{main_new}

\clearpage

\appendix{}

\onecolumngrid
\section{Numerical procedure}

\subsection{The norm constraint}
\begin{figure}[h!]
\centering
\includegraphics[width=0.95\textwidth]{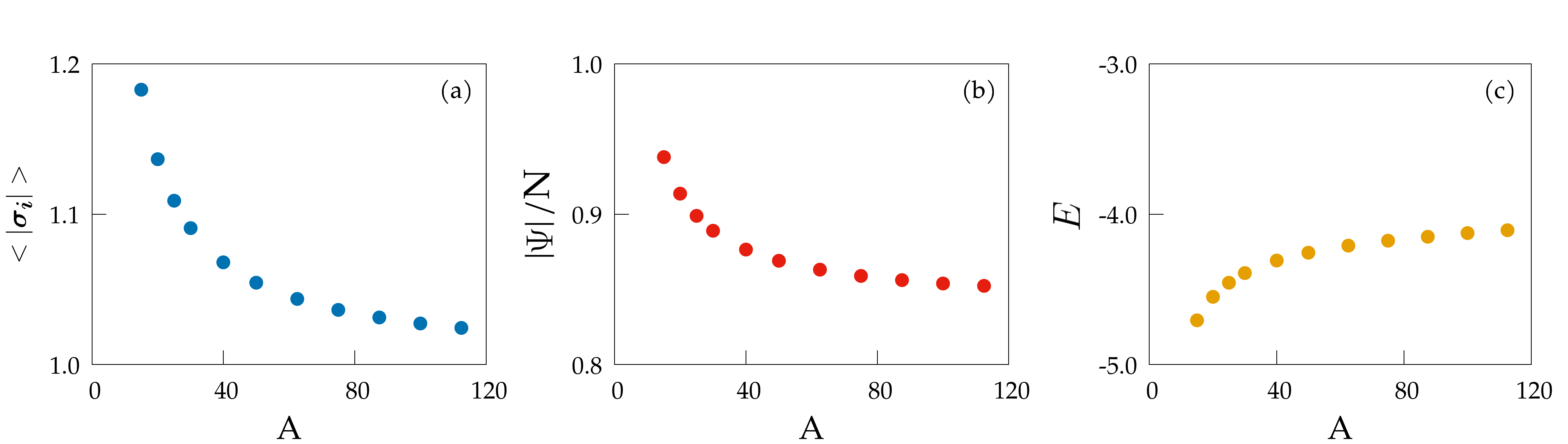}
\caption{a)  Mean norm $\langle\vert \boldsymbol{\sigma_i} \vert \rangle$  vs parameter $A$. b) Staggered magnetization per spin $\vert \boldsymbol{\Psi} \vert$  vs parameter $A$. c) Energy per spin  vs parameter $A$. The simulations have been done for $L=16$ , $\lambda=0.1$ and $T=0.6$.}
\label{fig_norm}
\end{figure}

The DLT equation of motion reads,
\begin{equation}
     \frac{d \boldsymbol{\sigma}_i}{dt} =
     \hbar^{-1} \frac{\partial H}{\partial \boldsymbol{\sigma}_i} \times \boldsymbol{\sigma}_i 
     - \hbar^{-1} \lambda \, \sum_{j}  \Lambda_{ij} \frac{\partial H}{\partial \boldsymbol{\sigma}_j}
     + \boldsymbol{\xi_i}.
     \label{eq:relax_azz}
\end{equation}
The first term (cross product) automatically conserves the norm of each spin. The second and third terms (relaxational and noise terms, respectively), however, would
produce some deviations from the initially fixed norm if no counter-measure is taken. In order to fix the norm of the spins one could use a Lagrange multiplier; however, 
the equation to work out the Lagrange multiplier would need to be solved numerically at each time step, hence slowing down significantly the simulation.
We therefore use a different method: we introduce a soft constraint on the norm by adding to the Hamiltonian the potential,
\begin{equation}
V_\mathrm{norm}= \frac{1}{4} A \sum_i\left(\boldsymbol{\sigma}_i^2 -1\right)^2  \ ,
\end{equation}
so that the spin force becomes,
\begin{equation}
  \frac{\partial H}{\partial \boldsymbol{\sigma}_i}=J\sum_j n_{ij} \boldsymbol{\sigma}_j + A \left(\boldsymbol{\sigma}_i^2 -1\right)\boldsymbol{\sigma}_i.
   \label{eq:Fnorma}
\end{equation}
It is easy to check that the reversible term is not affected by the soft constraint and that the total magnetization $\boldsymbol{M}$ remains constant. Hence, the soft constraint  does not modify the reversible dynamics and it conserves the total magnetization, independently of the value chosen for $A$. For $A \to \infty$, one would recover a strictly fixed norm, but of course for very large $A$ we would need a much higher precision to integrate the equation. To set $A$, therefore, we just need a large value that preserves in practice the norm, but does not demand an extreme precision in the simulations. Fortunately, because nothing essential in the dynamics actually depends on $A$, finding this balance is not too difficult.
We have fixed $A=100$ and we have carefully checked that this value does not play a relevant role in the results presented throughout the main text and the SI. In Fig. \ref{fig_norm} we can see that 
at this value of $A$, norm, energy, and order parameter are not far from their asymptotic, $A=\infty$, values.

\subsection{Identification of the scaling regime}

\begin{figure}[h!]
\centering
\includegraphics[width=0.75\textwidth]{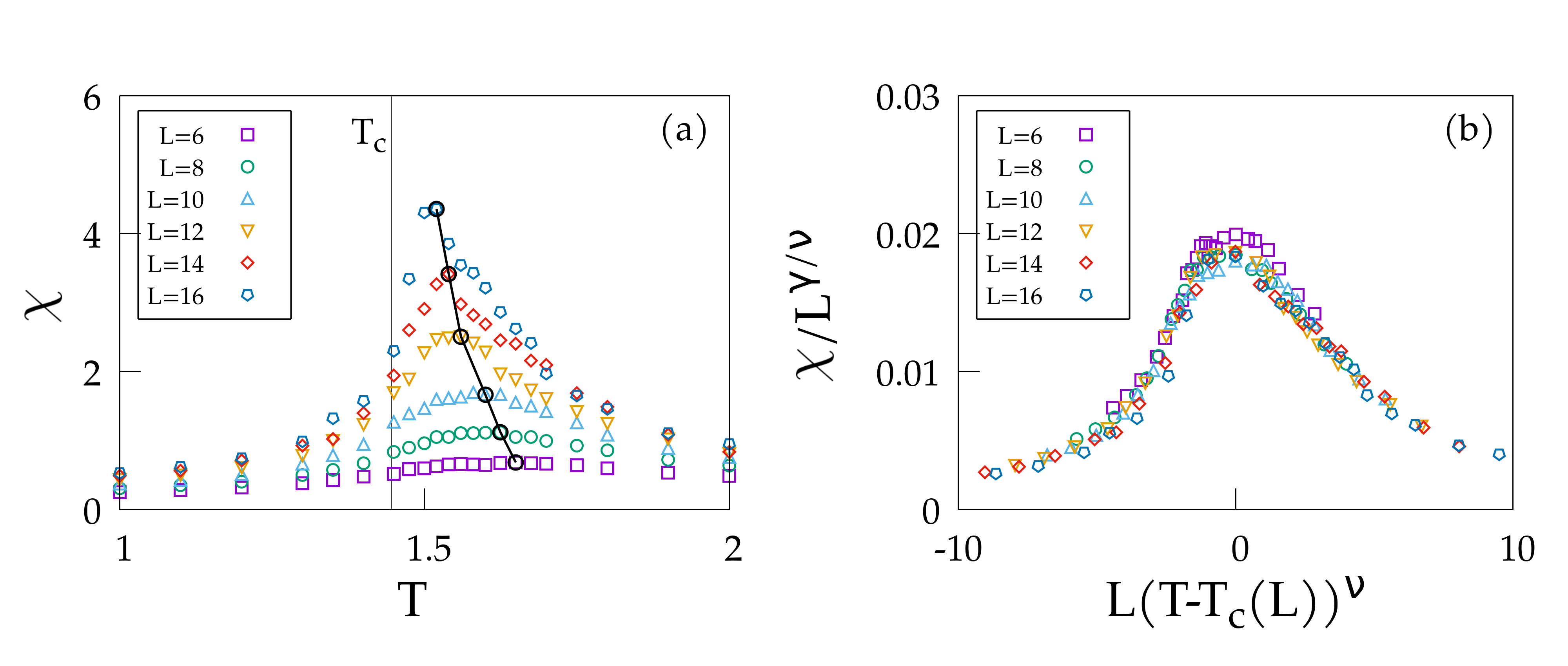}
\caption{a) Static susceptibility $\chi$ vs temperature $T$ for $\lambda=0.1$ and different system sizes $L$. The black circles correspond to the peak of the susceptibility for each size. The bulk critical temperature $T_c$ is marked by a vertical line. b) Collapse of the susceptibility $\chi$ by using the correct critical exponents.}
\label{fig_susc}
\end{figure}

The critical temperature of the Heisenberg antiferromagnet in $d=3$ is $T_c \sim 1.446J$ \cite{Tc_2001,Landau1993_1}. However, the actual phase transition only occurs in the thermodynamic limit, where $\xi\sim(T-T_c)^{-\nu} \to \infty$ and the static susceptibility $\chi \sim(T-T_c)^{-\gamma}  \to \infty$. In a finite system, the correlation length $\xi$ cannot diverge, as it is limited by the the system's size, $L$. 
Finite-size scaling \cite{privman1990finite} states that near-criticality the susceptibility $\chi$ (reported in Fig.\ref{fig_susc}a) obeys the relation,
\begin{equation}
    \chi = L^{\gamma/\nu} F\left(L(T-T_c)^\nu \right),
    \label{eq:sc}
\end{equation}
where $F(y)$ is a dimensionless scaling function. This scaling form implies that if we plot $\chi/L^{\gamma/\nu}$ vs $L(T-T_c)^\nu$, we should have a collapse of the data, provided that we use the correct critical exponents of the Heisenberg antiferromagnet in $d=3$ \cite{gammanu}, that is $\gamma/\nu = 1.97$ and $\nu=0.70$. This collapse is shown in Fig. \ref{fig_susc}b, which is already quite convincing. But to have a harder check, we need to work out the critical exponents in a more direct way than the collapse.

The most effective way to explore the critical point at finite size is to keep fixed the scaling variable, $y=L(T-T_c)^\nu$, thus defining a size-dependent pseudo-critical temperature, $T_c(L)$; the easiest way to do that is to locate the point $y_\mathrm{c}$ where $F(y)$ has its maximum, which corresponds to locating the temperature $T_c(L)$ where $\chi(T,L)$ has its maximum, at each given $L$ (see Fig.\eqref{fig_susc}a). In this scaling regime we have $\xi(T_c(L)) \sim L$, and
\begin{equation}
    \chi = L^{\gamma/\nu} F(y_c) \sim L^{\gamma/\nu} \ ,
    \label{pizza}
\end{equation}
because the scaling variable $y_c$ is kept constant by following the maximum of $\chi$ at each $L$. Equation \eqref{pizza} is tested in Fig.2d in the main text; the best fit to a log-log representation of the data gives $\gamma/\nu = 1.92 \pm 0.07$, quite close to the correct value.

\subsection{Relaxation time}

Given the dynamic correlation function of the order parameter in the momentum-frequency domain, $C(k,\omega)$, we can obtain the {\it static} correlation function, $C_0(k)$ as,
\begin{equation}
C_0(k) = \int_{-\infty}^\infty \frac{d\omega}{2\pi} \; C(k,\omega)
\end{equation}
which gives the following normalization condition, 
\begin{equation}
\int_{-\infty}^\infty \frac{d\omega}{2\pi} \; \frac{C(k,\omega)}{C_0(k)} = 1
\end{equation}
A convenient and very robust definition of the characteristic frequency $\omega_k$ is \cite{HH_scaling},
\begin{equation}
\int_{-\omega_k}^{\omega_k}\frac{d\omega}{2\pi}\;  \frac{C(k,\omega)}{C_0(k)}=\frac{1}{2}
\label{home}
\end{equation}
The characteristic frequency, $\omega_k$, is naturally the inverse of the relaxation time, $\tau_k$, which - working out \eqref{home} in the time domain - is defined by, 
\begin{equation}
    \int_0^{\infty} \frac{dt}{t} \;  \frac{C(k,t)}{C_0(k)}   \sin\left(t/\tau_k\right)  = \frac{\pi}{4}
\label{nope}
\end{equation}
which is the relation we use in our simulations to calculate the relaxation time (the transverse scattering function $S_\mathrm{T}(k,\omega)$ in the main text is the transverse part of $C(k,\omega)$). The advantage of calculating $\tau_k$ through equation \eqref{nope} is that no a priori fitting form for $C(k,t)$ is needed; moreover, being an integral equation that uses the entire temporal range of $C(k,t)$, this estimate is considerably more robust than crossing the correlation function with an arbitrary constant.

\subsection{Tests of thermalization}

\begin{figure}[h!]
\centering
\includegraphics[width=0.75\textwidth]{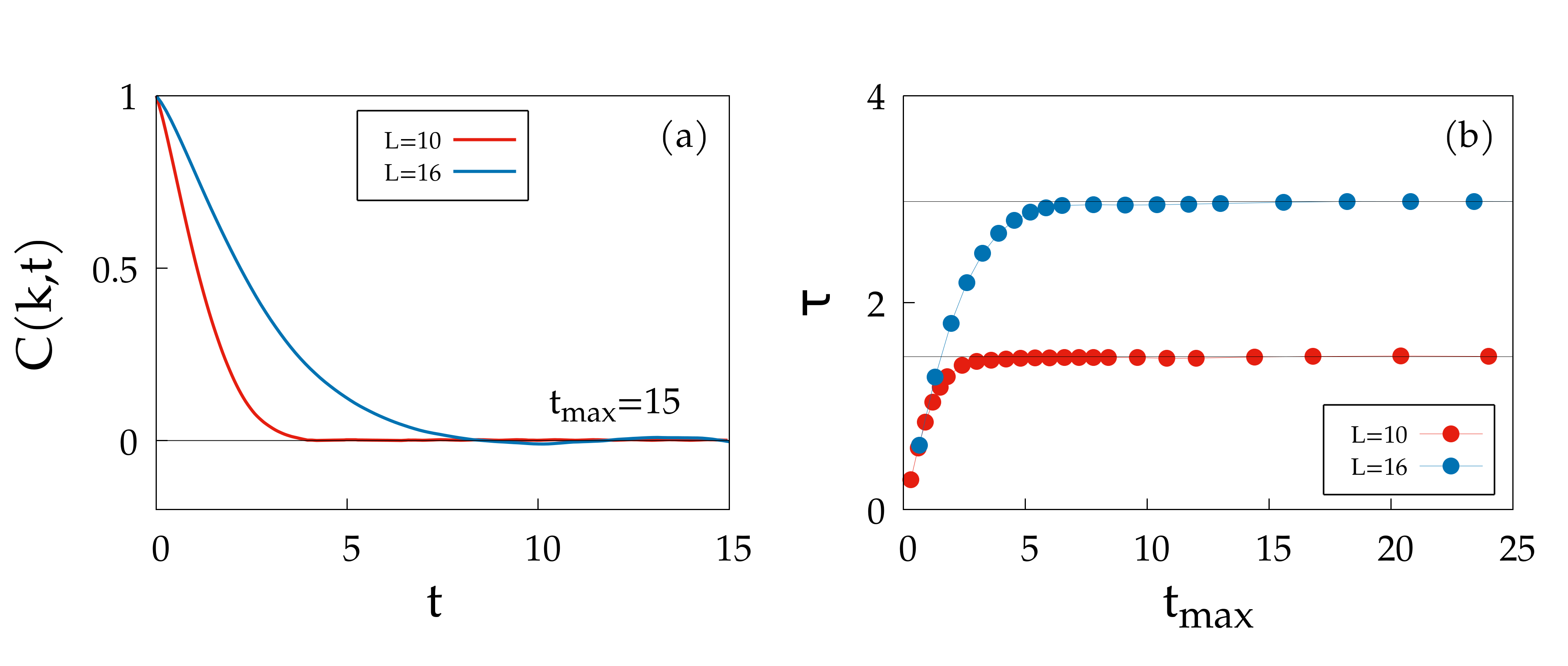}
\caption{a)  Correlation function $C(k,t)$ for two different system sizes $L$ ($\lambda=0.1$ and $T=T_c(L)$). b) Relaxation time $\tau$ vs trajectory duration $t_\mathrm{max}$. The lines correspond to the value of the stationary plateaus.}
\label{fig_ther}
\end{figure}

We test that the analysis to obtain the exponent $z$ has been done for a thermalized system. To do so, we check that the relaxation time $\tau$ does not depend on the duration of the simulation. We compute $\tau$ for trajectories of increasing duration $t_\mathrm{max}$ (see Fig. \eqref{fig_ther}). The value of $\tau$ grows with $t_\mathrm{max}$ until it reaches a stationary plateau. This procedure allows us to say that, when the duration is $t_\mathrm{max} \geq 10 \tau$, we are sure that the system has thermalized; we repeat this test for all temperatures $T$ and sizes $L$.

\clearpage

\section{Dispersion relation in the spin-wave regime}
In the main text, we report the dispersion relation for spin waves at low temperatures. In this section, we show how to derive this result for the case of the classical Heisenberg antiferromagnet in $d=3$.
Let us consider the microcanonical equation of motion,
\begin{equation}
      \dfrac{d\boldsymbol{\sigma_i}}{dt}= \dfrac{\partial H}{\partial \boldsymbol{\sigma_i}} \times \boldsymbol{\sigma_i},
      \label{sw_1}
\end{equation}
with $\dfrac{\partial H}{\partial \boldsymbol{\sigma_i}}=J \sum_j n_{ij} \boldsymbol{\sigma_j}$.
From \eqref{sw_1}, one obtains,
\begin{equation}
\begin{aligned}
      \dfrac{d^2\boldsymbol{\sigma_i}}{dt^2}=& 
     \dfrac{\partial H}{\partial \boldsymbol{\sigma_i}}\times\dfrac{d\boldsymbol{\sigma_i}}{dt}+\dfrac{d}{dt}\left(\dfrac{\partial H}{\partial \boldsymbol{\sigma_i}}\right) \times  \boldsymbol{\sigma_i}=\\
    =&  J \sum_j n_{ij} \left[\boldsymbol{\sigma_i} \times \dfrac{d\boldsymbol{\sigma_i}}{dt} +\dfrac{d\boldsymbol{\sigma_j}}{dt} \times \boldsymbol{\sigma_i}\right]=\\
    =& J \sum_j n_{ij} \left[\boldsymbol{\sigma_i} \times \left(\dfrac{\partial H}{\partial \boldsymbol{\sigma_i}} \times \boldsymbol{\sigma_i}\right)+ \left(\dfrac{\partial H}{\partial \boldsymbol{\sigma_j}} \times \boldsymbol{\sigma_i}\right)\right]=\\
    =& J^2 \sum_{jk} n_{ij} \left[\boldsymbol{\sigma_j} \times (n_{ik}\boldsymbol{\sigma_k} \times \boldsymbol{\sigma_i})+ (n_{jk}\boldsymbol{\sigma_k} \times \boldsymbol{\sigma_j}) \times \boldsymbol{\sigma_i}\right].
    \end{aligned}
    \label{sw_2}
\end{equation}
Multiplying both sides by the parity of the site $\pi_i=\pm 1$ (neighboring spins have opposite parity), we can rewrite Eq. \eqref{sw_2} in terms of the local staggered magnetizations, $\boldsymbol{\psi_i}=\pi_i\boldsymbol{\sigma_i}$,
\begin{equation}
      \dfrac{d^2\boldsymbol{\psi_i}}{dt^2}= J^2 \sum_{jk} n_{ij}\pi_j\pi_k \left[\boldsymbol{\psi_j} \times (n_{ik}\boldsymbol{\psi_k} \times \boldsymbol{\psi_i})+ (n_{jk}\boldsymbol{\psi_k} \times \boldsymbol{\psi_j}) \times \boldsymbol{\psi_i}\right].
      \label{eq:temp}
\end{equation}
In the first term of the r.h.s, the indexes  $j$ and $k$ have the same parity, while the opposite holds for the second term. Eq. \eqref{eq:temp} can then be rewritten as
 \begin{equation}
      \dfrac{d^2\boldsymbol{\psi_i}}{dt^2}= J^2 \sum_{jk} n_{ij} \left[\boldsymbol{\psi_j} \times (n_{ik}\boldsymbol{\psi_k} \times \boldsymbol{\psi_i})- (n_{jk}\boldsymbol{\psi_k} \times \boldsymbol{\psi_j}) \times \boldsymbol{\psi_i}\right] \ . 
\end{equation}
In the low temperature regime, i.e. $T \ll T_c$, the system is deeply polarized. We can thus expand the local staggered magnetization around the global polarization direction,
\beq
\boldsymbol{\hat{n}}=\frac{\sum_i \boldsymbol{\psi_i}}{\lvert \sum_i \boldsymbol{\psi_i} \rvert}
\eeq

that is, 
\beq
\boldsymbol{\psi_i}={\psi_i}^{\parallel} \boldsymbol{\hat{n}} +\boldsymbol{\pi_i}
\eeq 

with $\boldsymbol{\hat{n}}\cdot\boldsymbol{\pi_i}=0$, $\vert \boldsymbol{\pi_i} \vert \ll 1$, and where we have $\sum_i\boldsymbol{\pi_i}=0$. The spin wave expansion is an expansion in terms of the $\{\boldsymbol{\pi_i}\}$. Because of the constraint, $\vert \boldsymbol{\psi_i} \vert=1$, we have that ${\psi_i}^{\parallel}=\sqrt{1-\pi_i^2}$.  To first order, then, ${\psi_i}^{\parallel}\simeq 1$, and the equation of motion becomes,
\begin{equation}
\begin{aligned}
\dfrac{d^2\boldsymbol{\pi_i}}{dt^2}=&
J^2 \sum_{jk} n_{ij}[n_{ik} (\boldsymbol{\hat{n}} \times (\boldsymbol{\hat{n}} \times \boldsymbol{\pi}_i)+ \boldsymbol{\hat{n}} \times (\boldsymbol{\pi}_k \times \boldsymbol{\hat{n}}))-n_{jk} ((\boldsymbol{\hat{n}} \times \boldsymbol{\pi}_j) \times \boldsymbol{\hat{n}} + (\boldsymbol{\pi}_k \times \boldsymbol{\hat{n}}) \times \boldsymbol{\hat{n}}) ]\\
=&  J^2 \sum_{jk} [n_{ij}n_{ik} (-\boldsymbol{\pi_i}+\boldsymbol{\pi_k})-n_{ik}n_{jk} (-\boldsymbol{\pi_j}+\boldsymbol{\pi_k})] \ .
\end{aligned}
\end{equation}
If we now define (for a cubic lattice) $n_c=\sum_k n_{ik}=2d$, we get, 
\begin{equation}
\begin{aligned}
\dfrac{d^2\boldsymbol{\pi_i}}{dt^2}=& J^2  \left[-n_c^2 \boldsymbol{\pi_i} + n_c \sum_k n_{ik} \boldsymbol{\pi_k} + \sum_{jk} n_{ik}n_{jk}\boldsymbol{\pi_j}- n_c \sum_k n_{ik} \boldsymbol{\pi_k} \right]=\\
      =& J^2 \left[-n_c^2 \boldsymbol{\pi_i} +  \sum_{jk} n_{ik}n_{jk}\boldsymbol{\pi_j}\right]=\\
      =& J^2 \left[-n_c^2 \boldsymbol{\pi_i} +  \sum_{jk} (n_c \delta_{ik} -\Lambda_{ik})(\delta_{jk}n_c -\Lambda_{jk})\boldsymbol{\pi_j}\right] =\\
      =&  J^2 \sum_{jk} (\Lambda_{ik} \Lambda_{kj} - 2n_c \Lambda_{ij} \delta_{jk})\boldsymbol{\pi_i} / , 
\end{aligned}
\end{equation}
where, we remind, $\Lambda_{ij}=-n_{ij} + n_c \delta_{ij}$ is the discrete Laplacian. 
In Fourier space, this equation can be easily expressed in terms of the eigenvalues of the Laplacian operator
\begin{equation}
\lambda(\boldsymbol{q})=4 \sum_{\alpha=1}^d \sin^2\left(\frac{q_\alpha \ell}{2}\right),
\end{equation}The resulting dispersion relation reads,
\begin{equation}
 - \omega^2 (\boldsymbol{q})=  J^2 (\lambda^2(\boldsymbol{q})-2 n_c \lambda (\boldsymbol{q})).
\end{equation}
and taking $\boldsymbol{q}=(q,0,0)$, one obtains,
\begin{equation}
  \omega (q)=  4 J \sqrt{d} \sin \left( \frac{q \ell}{2} \right) \sqrt{1- \frac{1}{d} \sin^2 \left(\frac{q\ell}{2}\right)}.
  \label{sw_final}
\end{equation}
This expression is precisely the one reported in Eq. \eqref{eq:dis} and plotted in Fig.2f of the main text.
\clearpage


\section{Microscopic derivation of the Discrete Laplacian Thermostat}

\subsection{Hamiltonian of the coupled spin-lattice system}
Let us consider the magnon-phonon Hamiltonian, Eq. \eqref{H_tot_main} in the main text,
\begin{equation}
     H^{\mathrm{TOT}}=  \dfrac{1}{2}\sum_{ij} J(r_{ij}) n_{ij} \boldsymbol{\sigma}_i \cdot \boldsymbol{\sigma}_j + \frac{K}{2} \sum_{ij} n_{ij}(\boldsymbol{u}_i - \boldsymbol{u}_j)^2 + \frac{1}{2m} \sum_i \boldsymbol{p}_i^2  ,
\end{equation}\\
where, we remind,  $r_{ij}=|{\boldsymbol r}_i -{\boldsymbol r}_j|$ and  ${\boldsymbol u}_i={\boldsymbol r}_i - {\boldsymbol r}_i^{\rm 0}$ is the deviation of nucleus $i$ from its equilibrium position. 
\textcolor{black}{Following  \cite{soft_IV,Sanger1994,Xiong2017} we expand the spin exchange coupling to first order, 
\begin{equation}
     J(r_{ij})= J_0 \; n_{ij}   -\alpha \; n_{ij} \; (\boldsymbol{u}_i-\boldsymbol{u}_j) \cdot \hat{\boldsymbol{e}}_{ij}   \  , 
     \label{fartissimo}
\end{equation}
where,
\beq
J_0 = J(r_{ij}^0) \ ,
\eeq
is the spin-spin coupling constant at the leading order, while, 
\beq
\alpha = - J'(r_{ij}^0) \ , 
\eeq
is the spin-lattice coupling constant; $\alpha$ is positive, because the exchange interaction, $J(r)$, decreases with distance; we have also defined the normalized vector pointing from $j$ to $i$, 
\beq
\hat{\boldsymbol{e}}_{ij}=  (\boldsymbol{r}^0_{i}-\boldsymbol{r}^0_{j})/|\boldsymbol{r}^0_{i}-\boldsymbol{r}^0_{j}| \ .
\eeq
}
We can now rewrite the Hamiltonian as,
\begin{equation}
     H^{\mathrm{TOT}}=  \dfrac{J_0}{2}\sum_{ij}  n_{ij} \boldsymbol{\sigma}_i \cdot \boldsymbol{\sigma}_j - \alpha \sum_{ij}  n_{ij}\hat{\boldsymbol{e}}_{ij} \cdot (\boldsymbol{u}_i - \boldsymbol{u}_j) (\boldsymbol{\sigma}_i \cdot \boldsymbol{\sigma}_j) + \frac{K}{2} \sum_{ij} n_{ij}(\boldsymbol{u}_i - \boldsymbol{u}_j)^2 + \frac{1}{2m} \sum_i \boldsymbol{p}_i^2  \ .
\end{equation}
It is convenient to define the three distinct contributions to this Hamiltonian; the magnetic energy of the spins,
\begin{equation}
H_0=\dfrac{J_0}{2}\sum_{ij}  n_{ij} \boldsymbol{\sigma}_i \cdot \boldsymbol{\sigma}_j ,
\end{equation}
the vibrational energy of the phonons,
\begin{equation}
    H^{\mathrm{VIB}}= \frac{K}{2} \sum_{ij} n_{ij}(\boldsymbol{u}_i - \boldsymbol{u}_j)^2 + \frac{1}{2m} \sum_i \boldsymbol{p}_i^2 ,
\end{equation}
and finally the spin-lattice interaction energy,
\begin{equation}
    H^{\mathrm{INT}}= - \dfrac{\alpha}{2} \sum_{ij}  n_{ij}\hat{\boldsymbol{e}}_{ij} \cdot (\boldsymbol{u}_i - \boldsymbol{u}_j) (\boldsymbol{\sigma}_i \cdot \boldsymbol{\sigma}_j) .\\
\end{equation}
Periodic Boundary Conditions (PBC) are assumed, so that the system is translationally invariant.

Our aim in this section is to integrate out the contribution of phonons from the dynamics of the spin variables. To do so, it is convenient to rewrite the Hamiltonian in terms of variables that, contrary to the $\{{\boldsymbol u}_i\}$, are not coupled to each other.  Looking at the vibrational part, we  notice that the quadratic term can be rewritten as
\begin{equation}\label{laplacian_phononic_hamiltonian}
\frac{K}{2} \sum_{ij} n_{ij}(\boldsymbol{u}_i - \boldsymbol{u}_j)^2  =K \sum_{ij} \Lambda_{ij}\boldsymbol{u}_i\cdot \boldsymbol{u}_j ,
\end{equation}
where  $\Lambda_{ij}=-n_{ij}+\delta_{ij}\sum_{k} n_{ik}$ , is the discrete Laplacian.
The Hamiltonian \eqref{laplacian_phononic_hamiltonian} can then be rewritten in diagonal form using the eigenvectors and eigenvalues of the Laplacian
\begin{equation}
  \Lambda_{ij}w_j^{\boldsymbol{q}}=\lambda_{\boldsymbol{q}}w_i^{\boldsymbol{q}} ,  
\end{equation}
For a simple cubic lattice we have,
\begin{equation}
    w_i^{\boldsymbol{q}}= \frac{1}{\sqrt{N}}e^{-i\boldsymbol{q} \cdot \boldsymbol{r}_i} ,\quad \quad \lambda_{\boldsymbol{q}}=2\sum_{\alpha=1}^d(1- \cos(q^{\alpha} \ell) )
\end{equation}
with $\ell$ the lattice spacing. Due to the PBC, $\boldsymbol{q}$ can only assume discrete values, $\boldsymbol{q}=\frac{2\pi}{L}(n_x, n_y, n_z)$, with $L$ the lattice size, $n_x$, $n_y$, $n_z$ integers from 0 to $n_{max}=(L/\ell -1)$.
The change of basis using the eigenvectors of the discrete Laplacian gives,
\begin{equation}
\begin{aligned}
    \boldsymbol{u}_i=&\sum_{\boldsymbol{q}} \boldsymbol{Q}_{\boldsymbol{q}}\frac{1}{\sqrt{N}}e^{-i\boldsymbol{q} \cdot \boldsymbol{r}_i} , \quad \quad 
    \boldsymbol{Q}_{\boldsymbol{q}}=&\sum_{i}\boldsymbol{u}_i \frac{1}{\sqrt{N}}e^{+i\boldsymbol{q} \cdot \boldsymbol{r}_i} , \\ 
    \boldsymbol{p}_i=&\sum_{\boldsymbol{q}} \boldsymbol{P}_{\boldsymbol{q}}\frac{1}{\sqrt{N}}e^{-i\boldsymbol{q} \cdot \boldsymbol{r}_i} , \quad \quad
    \boldsymbol{P}_{\boldsymbol{q}}=&\sum_{i}\boldsymbol{p}_i \frac{1}{\sqrt{N}}e^{+i\boldsymbol{q} \cdot \boldsymbol{r}_i} , \\
\end{aligned}
\end{equation}
and
\begin{equation}
    \sum_{ij} w_i^{-\boldsymbol{q}} \Lambda_{ij} w_j^{\boldsymbol{p}} = \delta_{\boldsymbol{q}\boldsymbol{p}} \lambda_{\boldsymbol{q}} .
\end{equation}\\
Hence we have, 
\begin{equation}
    H^{\mathrm{VIB}}= K \sum_{\boldsymbol{q}} \lambda_{\boldsymbol{q}} \boldsymbol{Q}_{\boldsymbol{q}} \boldsymbol{Q}_{-\boldsymbol{q}} + \sum_{\boldsymbol{q}} \frac{1}{2m} \boldsymbol{P}_{\boldsymbol{q}} \cdot \boldsymbol{P}_{-\boldsymbol{q}};
\end{equation}
To rewrite the interaction energy we define the auxiliary quantities, 
\begin{equation}\label{eq_f}
    \begin{aligned}
      \boldsymbol{f}_{i}(\{\boldsymbol{\sigma}\})=&  \sum_j n_{ij}\hat{\boldsymbol{e}}_{ij} (\boldsymbol{\sigma}_i \cdot \boldsymbol{\sigma}_j) ,\\
\boldsymbol{f}_{\boldsymbol{q}}(\{\boldsymbol{\sigma}\})=& \sum_i \frac{1}{\sqrt{N}}e^{i\boldsymbol{q} \cdot \boldsymbol{r}_i} \boldsymbol{f}_{i}(\{\boldsymbol{\sigma}\}) ,
    \end{aligned}
\end{equation}
and so we get,
\begin{equation}
\begin{aligned}
    H^{\mathrm{INT}}=& - \dfrac{\alpha}{2} \sum_{ij}  n_{ij}\hat{\boldsymbol{e}}_{ij} \cdot (\boldsymbol{u}_i - \boldsymbol{u}_j) (\boldsymbol{\sigma}_i \cdot \boldsymbol{\sigma}_j) =\\
    =& - \dfrac{\alpha}{2} \sum_{ij}  n_{ij}\hat{\boldsymbol{e}}_{ij} \cdot \boldsymbol{u}_i (\boldsymbol{\sigma}_i \cdot \boldsymbol{\sigma}_j) + \dfrac{\alpha}{2} \sum_{ij}  n_{ij}\hat{\boldsymbol{e}}_{ij} \cdot \boldsymbol{u}_j (\boldsymbol{\sigma}_i \cdot \boldsymbol{\sigma}_j)=\\
    =& -\alpha \sum_{ij}  n_{ij}\hat{\boldsymbol{e}}_{ij} \cdot \boldsymbol{u}_i (\boldsymbol{\sigma}_i \cdot \boldsymbol{\sigma}_j)= - \alpha \sum_{i}  \boldsymbol{u}_{i} \cdot \boldsymbol{f}_{i}(\{\boldsymbol{\sigma}\}) ;
    \end{aligned}
\end{equation}       
hence,
\begin{equation}
    H^{\mathrm{INT}}= -\alpha \sum_{\boldsymbol{q}} \boldsymbol{Q}_{\boldsymbol{q}} \cdot \boldsymbol{f}_{\boldsymbol{-q}}(\{\boldsymbol{\sigma}\})=-\alpha \sum_{\boldsymbol{q}} \boldsymbol{Q}_{-\boldsymbol{q}} \cdot \boldsymbol{f}_{\boldsymbol{q}}(\{\boldsymbol{\sigma}\}) .
\end{equation}
Finally, combining all terms, we end up with,      
\begin{equation}\label{H_tot}
    H^{\mathrm{TOT}}=J_0 \sum_{ij} n_{ij} (\boldsymbol{\sigma}_i \cdot \boldsymbol{\sigma}_j) -\alpha \sum_{\boldsymbol{q}} \boldsymbol{Q}_{\boldsymbol{q}} \cdot \boldsymbol{f}_{\boldsymbol{-q}}(\{\boldsymbol{\sigma}\}) +  K \sum_{\boldsymbol{q}} \lambda_{\boldsymbol{q}} \boldsymbol{Q}_{\boldsymbol{q}} \boldsymbol{Q}_{-\boldsymbol{q}}  +  \sum_{\boldsymbol{q}} \frac{1}{2m} \boldsymbol{P}_{\boldsymbol{q}} \cdot \boldsymbol{P}_{-\boldsymbol{q}} .
\end{equation}
We will now write the equations of motion for both the spins and the lattice vibrational degrees of freedom. In the new basis, the variables $\{\boldsymbol{Q}_{\boldsymbol{q}}\}$ are not directly coupled one with the other; we can therefore solve explicitly their equations and plug back the solutions in the equations for the spins. The result is an equation of motion for the spin variables only, where the effect of the phonons is  the appearance of a thermal bath composed by a "noise" and a "relaxational"  term, both of which preserve the conservation of the global magnetization. This thermal bath always satisfies the Fluctuation Dissipation (FD) relation, but the specific memory kernel will depend on the details of the dynamics. In the following sections, we are going to derive it for both an isolated system, and in the case where the nuclei, but not the spins, are in contact with a (standard) thermal bath.
     
     
\subsection{Isolated system}\label{sec:microcanonical}
\subsubsection{Equations of motion}\label{sec:eq_motion_micro}
The outline for deriving a thermal bath out of many oscillatory degrees of freedom of an isolated system can be found in \cite{Zwanzig}; in the case of ferromagnetic systems, a similar computation has been done in \cite{Rossi_MacDonald} within a mesoscopic field theory framework. We start by writing the equations of motion for phonons and spins, 
\begin{equation}\label{eq_of_motion_phonons_micro}
    m \dfrac{d^2\boldsymbol{Q}_{\boldsymbol{q}}}{dt^2}=\dfrac{d\boldsymbol{P}_{\boldsymbol{q}}}{dt} =-\dfrac{\partial H^{\mathrm{TOT}}}{\partial \boldsymbol{Q}_{-\boldsymbol{q}}} = - 2K \lambda_{\boldsymbol{q}} \boldsymbol{Q}_{\boldsymbol{q}} + \alpha \boldsymbol{f}_{\boldsymbol{q}} ,
\end{equation}
\begin{equation}\label{eq_motion_spins_H_TOT}
    \dfrac{d \boldsymbol{\sigma}_i}{dt} =  \hbar^{-1} \dfrac{\partial H^{\mathrm{TOT}}}{\partial \boldsymbol{\sigma}_i} \times \boldsymbol{\sigma}_i=\hbar^{-1} \dfrac{\partial H_0}{\partial \boldsymbol{\sigma}_i} \times \boldsymbol{\sigma}_i - \dfrac{\alpha}{\hbar} \sum_{\boldsymbol{q},\beta}  Q^{\beta}_{\boldsymbol{q}} \dfrac{\partial  f^{\beta}_{-\boldsymbol{q}}(\{\boldsymbol{\sigma} \}) }{\partial \boldsymbol{\sigma}_i} \times \boldsymbol{\sigma}_i .\\
    \end{equation}
The plan is now to solve Eq.\eqref{eq_of_motion_phonons_micro} and plug the result back  into Eq.\eqref{eq_motion_spins_H_TOT}. The solution of the homogeneous equation is given by,
\begin{equation}
\boldsymbol{Q}^{\mathrm{HOM}}_{\boldsymbol{q}}(t) = \boldsymbol{Q}_{\boldsymbol{q}}(0) \cos(\omega_{\boldsymbol{q}}t) + \dfrac{\boldsymbol{P}_{\boldsymbol{q}}(0)}{m \omega_{\boldsymbol{q}}} \sin(\omega_{\boldsymbol{q}}t) ,
\end{equation}
where $\boldsymbol{Q}_{\boldsymbol{q}}(0)$ and $\boldsymbol{P}_{\boldsymbol{q}}(0)$ are the initial conditions at time $t=t_0=0$, and,
\begin{equation}
    \omega^2_{\boldsymbol{q}}= \dfrac{2K}{m} \lambda_{\boldsymbol{q}} .
\end{equation}
The complete solution is therefore given by,
\begin{equation}
\begin{aligned}
\boldsymbol{Q}^{\mathrm{TOT}}_{\boldsymbol{q}}(t) = & \boldsymbol{Q}^{\mathrm{HOM}}_{\boldsymbol{q}}(t) + \alpha \int_{0}^t dt' \dfrac{\sin(\omega_{\boldsymbol{q}}(t-t'))}{m \omega_{\boldsymbol{q}}} \boldsymbol{f}_{\boldsymbol{q}}(t')=\\
= & \boldsymbol{Q}^{\mathrm{HOM}}_{\boldsymbol{q}}(t) + \dfrac{\alpha}{m\omega_{\boldsymbol{q}}^2}\boldsymbol{f}_{\boldsymbol{q}}(t) - \dfrac{\alpha}{m\omega_{\boldsymbol{q}}^2}\boldsymbol{f}_{\boldsymbol{q}}(0) \cos(\omega_{\boldsymbol{q}}(t) ) - \dfrac{\alpha}{m \omega_{\boldsymbol{q}}^2} \int_{0}^t dt' \cos(\omega_{\boldsymbol{q}}(t-t')) \dfrac{d}{dt'}\boldsymbol{f}_{\boldsymbol{q}}(t') \ .
\end{aligned}
\end{equation}
Substituting into the equations of motion for the spins we get,
\begin{equation}\label{eq_motion_spin_thermostat}
    \dfrac{d \boldsymbol{\sigma}_i}{dt} = \hbar^{-1} \dfrac{\partial H_{\mathrm{eff}}}{\partial \boldsymbol{\sigma}_i} \times \boldsymbol{\sigma}_i - \boldsymbol{\Xi}_i\left[\{\boldsymbol{\sigma}\}\right] +\boldsymbol{\xi}_i ,
\end{equation}
where the effective Hamiltonian is given by,
\begin{equation}\label{def_H_eff}
   H_{\mathrm{eff}}= H_0 -  \sum_{\boldsymbol{q}}  \frac{\alpha^2}{2m\omega_{\boldsymbol{q}}^2}  \boldsymbol{f}_{\boldsymbol{q}} \cdot  \boldsymbol{f}_{-\boldsymbol{q}}(\{\boldsymbol{\sigma}\}) , 
\end{equation}
the relaxational term is given by (square brackets indicate functional dependence), 
\begin{equation}\label{def_relaxational}
    \boldsymbol{\Xi}_i\left[\{\boldsymbol{\sigma}\}\right] = - \sum_{\boldsymbol{q}\beta}   \dfrac{\alpha^2}{m \hbar \omega_{\boldsymbol{q}}^2} \int_{0}^t dt' \cos(\omega_{\boldsymbol{q}}(t-t')) \dfrac{d}{dt'}f^{\beta}_{\boldsymbol{q}}(t') \left(\dfrac{\partial   f^{\beta}_{-\boldsymbol{q}}(\{\boldsymbol{\sigma} \}) }{\partial \boldsymbol{\sigma}_i} \times \boldsymbol{\sigma}_i \right)_t ,
\end{equation}
and the noise term is given by,
\begin{equation}\label{def_noise}
  \boldsymbol{\xi}_i = -\dfrac{\alpha}{\hbar} \sum_{\boldsymbol{q}\beta} \left( Q^{\beta}_{\boldsymbol{q}}(0) \cos(\omega_{\boldsymbol{q}}t) + \dfrac{P^{\beta}_{\boldsymbol{q}}(0)}{m\omega_{\boldsymbol{q}}} \sin(\omega_{\boldsymbol{q}}t) - \dfrac{\alpha}{m\omega_{\boldsymbol{q}}^2}f^{\beta}_{\boldsymbol{q}}(0) \cos(\omega_{\boldsymbol{q}}t) \right) \dfrac{\partial   f^{\beta}_{-\boldsymbol{q}}(\{\boldsymbol{\sigma} \}) }{\partial \boldsymbol{\sigma}_i} \times \boldsymbol{\sigma}_i \ .
\end{equation}
Equation \eqref{eq_motion_spins_H_TOT} is the same equation of motion (written in Fourier's space) as Eq. \eqref{suka} of the main text, used to find the dependence of $\lambda$ on the microscopic parameters. We notice, however, that the spin-lattice interaction term in the r.h.s. of Eq. \eqref{eq_motion_spins_H_TOT}, which depends on $\boldsymbol{Q}_{\boldsymbol{q}}$, is generating not only the noise (as in the simplified argument of the main text), but also the relaxational term. This is what we should expect, since  they always emerge together in actual physical systems.


\subsubsection{Noise}
In the context of the microcanonical dynamics discussed in this section, if we want to describe the properties of the system at a given temperature $T$, the natural choice is to consider the initial conditions as drawn with a canonical distribution $e^{-\beta H}/Z$. We note that the total Hamiltonian of  Eq. \eqref{H_tot} can be rewritten as,
\begin{equation}
    H^{\mathrm{TOT}}=H_{\mathrm{eff}}(\{ \boldsymbol{\sigma}\}) + \sum_{\boldsymbol{q}}  \dfrac{m}{2}  \left(\omega_{\boldsymbol{q}} \boldsymbol{Q}_{\boldsymbol{q}} -\dfrac{\alpha}{m\omega_{\boldsymbol{q}}}  \boldsymbol{f}_{\boldsymbol{q}} \right) \cdot \left(\omega_{\boldsymbol{q}} \boldsymbol{Q}_{-\boldsymbol{q}} -\dfrac{\alpha}{m\omega_{\boldsymbol{q}}}  \boldsymbol{f}_{-\boldsymbol{q}} \right) + \sum_{\boldsymbol{q}} \frac{1}{2m} \boldsymbol{P}_{\boldsymbol{q}} \cdot \boldsymbol{P}_{-\boldsymbol{q}} .
\end{equation}       
We therefore have for the canonical averages,
\begin{equation}
\langle Q^{\alpha}_{\boldsymbol{q}}(0) - \dfrac{\alpha}{m\omega^2_{\boldsymbol{q}}}f^{\alpha}_{\boldsymbol{q}}(0) \rangle =0 ,
\label{average_Q}
\end{equation}
\begin{equation}\label{equipartizione_Q}
\dfrac{m\omega^2_{\boldsymbol{q}}}{2} \bigg \langle \bigg( Q^{\alpha}_{\boldsymbol{q}}(0) - \dfrac{\alpha}{m\omega^2_{\boldsymbol{q}}} f^{\alpha}_{\boldsymbol{q}}(0) \bigg) \bigg(  Q^{\beta}_{\boldsymbol{q}'}(0) - \dfrac{\alpha}{m\omega^2_{\boldsymbol{q}'}}f^{\beta}_{\boldsymbol{q}'}(0) \bigg) \bigg\rangle = \dfrac{k_BT}{2} \delta_{\alpha, \beta} \delta_{q',-q} ,
\end{equation}
\begin{equation}
\langle P^{\alpha}_{\boldsymbol{q}}(0)\rangle =0 ,
\end{equation}
\begin{equation}\label{equipartizione_P}
\dfrac{1}{2m}\langle P^{\alpha}_{\boldsymbol{q}}(0) P^{\beta}_{\boldsymbol{q}'}(0) \rangle = \dfrac{k_BT}{2}\delta_{\alpha, \beta} \delta_{q',-q} .
\end{equation}
We can exploit these expressions to compute the properties of the effective noise $\boldsymbol{\xi}_i $. From Eq.\eqref{def_noise}) we get,
\begin{equation}
    \langle \boldsymbol{\xi}_i \rangle = 0,
\end{equation}
and, from Eqs. \eqref{equipartizione_Q} and \eqref{equipartizione_P}, 
\begin{align}
     \langle \xi^{\mu}_i(t) \xi^{\nu}_j(t') \rangle =& \dfrac{\alpha^2}{\hbar^2} \sum_{\alpha\beta\boldsymbol{q}\boldsymbol{q}'} \bigg[ 
    \bigg  \langle \bigg( Q^{\alpha}_{\boldsymbol{q}}(0) - \dfrac{\alpha}{m\omega^2_{\boldsymbol{q}}} f^{\alpha}_{\boldsymbol{q}}(0) \bigg) \bigg(  Q^{\beta}_{\boldsymbol{q}'}(0) - \dfrac{\alpha}{m\omega^2_{\boldsymbol{q}'}}f^{\beta}_{\boldsymbol{q}'}(0) \bigg) \bigg \rangle \cos(\omega_{\boldsymbol{q}}t)\cos(\omega_{\boldsymbol{q}'}t')\  + \quad \quad \quad \quad \quad \quad \quad \quad \nonumber   \\
   +& \dfrac{\langle P^{\alpha}_{\boldsymbol{q}}(0) P^{\beta}_{\boldsymbol{q}'}(0) \rangle}{m^2\omega_{\boldsymbol{q}}\omega_{\boldsymbol{q}'}} \sin(\omega_{\boldsymbol{q}}t) \sin(\omega_{\boldsymbol{q}'}t') \bigg] \bigg(  \dfrac{\partial f^{\alpha}_{-\boldsymbol{q}}(\{\boldsymbol{\sigma} \})}{\partial \boldsymbol{\sigma}_i} \times \boldsymbol{\sigma}_i \bigg)^{\mu}_{t} \bigg(  \dfrac{\partial f^{\beta}_{-\boldsymbol{q}'}(\{\boldsymbol{\sigma} \}) }{\partial \boldsymbol{\sigma}_j} \times  \boldsymbol{\sigma}_j \bigg)^{\nu}_{t'}= \nonumber \\
    =& \sum_{\beta \boldsymbol{q}} \dfrac{\alpha^2 k_B T}{m \hbar^2 \omega^2_{\boldsymbol{q}}} \bigg[ \cos(\omega_{\boldsymbol{q}}t)\cos(\omega_{\boldsymbol{q}'}t') + \sin(\omega_{\boldsymbol{q}}t) \sin(\omega_{\boldsymbol{q}}t') \bigg]   \bigg(  \dfrac{\partial f^{\beta}_{-\boldsymbol{q}}(\{\boldsymbol{\sigma} \})}{\partial \boldsymbol{\sigma}_i} \times \boldsymbol{\sigma}_i \bigg)^{\mu}_t \bigg(  \dfrac{\partial f^{\beta}_{\boldsymbol{q}}(\{\boldsymbol{\sigma} \}) }{\partial \boldsymbol{\sigma}_j} \times  \boldsymbol{\sigma}_j \bigg)^{\nu}_{t'}= \nonumber \\ 
     = & \sum_{\beta\boldsymbol{q}} \dfrac{\alpha^2}{2\hbar^2 K\lambda_{\boldsymbol{q}}} \cos(\omega_{\boldsymbol{q}}(t-t')) \bigg(  \dfrac{\partial f^{\beta}_{-\boldsymbol{q}}(\{\boldsymbol{\sigma} \})}{\partial \boldsymbol{\sigma}_i} \times \boldsymbol{\sigma}_i \bigg)^{\mu}_t \bigg(  \dfrac{\partial f^{\beta}_{\boldsymbol{q}}(\{\boldsymbol{\sigma} \}) }{\partial \boldsymbol{\sigma}_j} \times  \boldsymbol{\sigma}_j \bigg)^{\nu}_{t'} .
     \label{eq:noise_kernel_micro}
\end{align}
Hence, the $\{\boldsymbol{\xi}_i\}$ are random variables with a Gaussian distribution, zero mean and non-trivial correlations both in space, as noises in different sites $i$ and $j$ are not independent, and in time.  We can indeed identify in Eq. \eqref{eq:noise_kernel_micro} the following non-Markovian, dimensionless memory kernel,
\begin{equation}\label{dimensionless_Kernel_micro}
{\cal R}^{\mu\nu}_{ij}(t,t')   = \sum_{\beta\boldsymbol{q}} \frac{1}{4\lambda_{\boldsymbol{q}}} \cos(\omega_{\boldsymbol{q}}(t-t')) \bigg(  \dfrac{\partial f^{\beta}_{-\boldsymbol{q}}(\{\boldsymbol{\sigma} \})}{\partial \boldsymbol{\sigma}_i} \times \boldsymbol{\sigma}_i \bigg)^{\mu}_t \bigg(  \dfrac{\partial f^{\beta}_{\boldsymbol{q}}(\{\boldsymbol{\sigma} \}) }{\partial \boldsymbol{\sigma}_j} \times  \boldsymbol{\sigma}_j \bigg)^{\nu}_{t'}  \ .
\end{equation}
The noise correlator can then be expressed in a more compact form as,
\begin{equation}\label{eq:correlation_noise_kernel}
    \langle \xi^{\mu}_i(t) \xi^{\nu}_j(t') \rangle = 2k_BT \hbar^{-1} \left(\dfrac{\alpha^2}{\hbar K}\right) {\cal R}^{\mu\nu}_{ij}(t,t') .
\end{equation}
Equation \eqref{eq:correlation_noise_kernel} is the same eq. as \eqref{sandapa} of the main text, and already looks quite  similar to \eqref{sandapapero}, written in the main text using dimensional analysis. Some more work, however, is needed to recover the same thermostat as DLT.

\subsubsection{Relaxational term and the FD relation}   

We now consider the relaxational term $\boldsymbol{\Xi}_i\left[\{\boldsymbol{\sigma}\}\right]$ of Eq. \eqref{def_relaxational}. We can exploit Eq. \eqref{eq_motion_spins_H_TOT} to expand the time derivative of $f^{\beta}_{\boldsymbol{q}}$, 
\begin{equation}\label{eq:time_derivative_f}
    \dfrac{d f^{\beta}_{\boldsymbol{q}}}{dt}= \sum_{j} \dfrac{\partial f^{\beta}_{\boldsymbol{q}}}{\partial \boldsymbol{\sigma}_j} \cdot \dfrac{d\boldsymbol{\sigma}_j}{dt'} = \dfrac{\partial f^{\beta}_{\boldsymbol{q}}}{\partial\boldsymbol{\sigma}_j} \cdot \left( \hbar^{-1}\dfrac{\partial H^{\mathrm{TOT}}}{\partial \boldsymbol{\sigma}_j} \times \boldsymbol{\sigma}_j \right) = -\hbar^{-1} \dfrac{\partial H^{\mathrm{TOT}}}{\partial \boldsymbol{\sigma}_j} \cdot \left( \dfrac{\partial f^{\beta}_{\boldsymbol{q}}}{\partial\boldsymbol{\sigma}_j} \times \boldsymbol{\sigma}_j \right) .
\end{equation}
Using the above equation we find,
\begin{equation}\label{eq:friction_kernel_micro}
\begin{aligned}
\boldsymbol{\Xi}_i\left[\{\boldsymbol{\sigma}\}\right] = & \sum_{j \boldsymbol{q}, \beta}   \dfrac{\alpha^2}{m\hbar \omega_{\boldsymbol{q}}^2} \int_0^t dt' \cos(\omega_{\boldsymbol{q}}(t-t')) \left( \dfrac{\partial   f^{\beta}_{-\boldsymbol{q}}(\{\boldsymbol{\sigma} \}) }{\partial \boldsymbol{\sigma}_i} \times \boldsymbol{\sigma}_i \right)_t \left( \dfrac{\partial f^{\beta}_{\boldsymbol{q}}}{\partial\boldsymbol{\sigma}_j} \times \boldsymbol{\sigma}_j \right)_{t'} \cdot \left( \hbar^{-1} \dfrac{\partial H^{\mathrm{TOT}}}{\partial \boldsymbol{\sigma}_j} \right)_{t'} .\\
\end{aligned}
\end{equation}
We then find that the relaxational term can be written as,
\begin{equation}\label{relaxational_kernel_micro_indices}
  \Xi_i^{\mu}=\hbar^{-1} \left(\dfrac{\alpha^2}{\hbar K}\right)\int_0^t dt' \sum_{j,\nu}  2   {\cal R}^{\mu\nu}_{ij}(t,t') \left( \dfrac{\partial H^{\mathrm{TOT}}}{\partial \sigma^{\nu}_j} \right)_{t'} ,
\end{equation}
Hence, the same memory kernel appears in the relaxational term and in the noise correlator, Eq.\eqref{eq:correlation_noise_kernel}. Together with the factor $k_BT$, this fully recovers the FD relation. We recall here that the FD relation must be satisfied in order to have a thermal bath, which preserves the equilibrium probability distribution of the system.

\subsubsection{Conservation of the total Magnetization}

Let us prove here that the dimensionless kernel appearing in both the noise and relaxational terms preserves the total magnetization along the dynamics. Let us consider Eq. \eqref{eq_motion_spin_thermostat}, the sum over $i$ yields the derivative of the total magnetization of the system,
\begin{equation}
    \dfrac{d \boldsymbol{M}}{dt} = \hbar^{-1}\sum_i \left ( \dfrac{\partial H_{\mathrm{eff}}}{\partial \boldsymbol{\sigma}_i} \times \boldsymbol{\sigma}_i \right )- \sum_i \boldsymbol{\Xi}_i\left[\{\boldsymbol{\sigma}\}\right] +\sum_i \boldsymbol{\xi}_i .
    \label{eq:mag}
\end{equation}
We now show that each term of the r.h.s. is zero. Recalling the definition of $f^\beta_{\boldsymbol{\sigma}}$ in Eq \eqref{eq_f}, we note that,
\begin{equation}
    \begin{aligned}
        \dfrac{\partial f^{\beta}_{-\boldsymbol{q}}}{\partial\boldsymbol{\sigma}_i} =& \sum_k \frac{1}{\sqrt{N}}e^{-i\boldsymbol{q} \cdot \boldsymbol{r}_k} \dfrac{ \partial f^{\beta}_{k}(\{\boldsymbol{\sigma}\})}{\partial \boldsymbol{\sigma}_i}= \sum_k \frac{1}{\sqrt{N}}e^{-i\boldsymbol{q} \cdot \boldsymbol{r}_k} \sum_h  n_{ij}\ \hat{\boldsymbol{e}}_{ij}^{\beta} \ \dfrac{\partial( \boldsymbol{\sigma}_k \cdot \boldsymbol{\sigma}_h)}{\partial \boldsymbol{\sigma}_i}=\\
        =& \sum_k \frac{1}{\sqrt{N}}e^{-i\boldsymbol{q} \cdot \boldsymbol{r}_k}  \sum_h n_{ij}\ \hat{\boldsymbol{e}}_{ij}^{\beta} \ \dfrac{\partial( \boldsymbol{\sigma}_k \cdot \boldsymbol{\sigma}_h)}{\partial \boldsymbol{\sigma}_i}= \\
        =& \sum_k \frac{1}{\sqrt{N}}e^{-i\boldsymbol{q} \cdot \boldsymbol{r}_k}  \sum_h n_{ij}\ \hat{\boldsymbol{e}}_{ij}^{\beta}\ 
        \left[ \delta_{ik} \boldsymbol{\sigma}_h + \delta_{ih} \boldsymbol{\sigma}_k  \right]= \\
        =& \sum_{kh} \frac{1}{\sqrt{N}} \left(e^{-i\boldsymbol{q} \cdot \boldsymbol{r}_k} -  e^{i\boldsymbol{q} \cdot \boldsymbol{r}_h} \right)  n_{ij}\ \hat{\boldsymbol{e}}_{ij}^{\beta}\ \delta_{ik} \boldsymbol{\sigma}_h  = \\
        =& \sum_{h} \frac{1}{\sqrt{N}} \left(e^{-i\boldsymbol{q} \cdot \boldsymbol{r}_i} -  e^{-i\boldsymbol{q} \cdot \boldsymbol{r}_h} \right)  n_{ij} \ \hat{\boldsymbol{e}}_{ij}^{\beta} \ \boldsymbol{\sigma}_h \ .
    \end{aligned}
\end{equation}
From this expression we get,
\begin{equation}\label{eq_half_pseudo_laplacian}
 \sum_i  \left(\dfrac{\partial f^{\beta}_{\boldsymbol{-q}}}{\partial\boldsymbol{\sigma}_i} \times \boldsymbol{\sigma}_i \right) = \sum_{i h} \frac{1}{\sqrt{N}} \left(e^{-i\boldsymbol{q} \cdot \boldsymbol{r}_i} -  e^{-i\boldsymbol{q} \cdot \boldsymbol{r}_h} \right)  n_{ij} \ \hat{\boldsymbol{e}}_{ij}^{\beta} \ ( \boldsymbol{\sigma}_h \times \boldsymbol{\sigma}_i) = 0 \ ,
\end{equation}
where the last passage derives from summing an anti-symmetric object. \textcolor{black}{Thanks to this relation, and recalling \eqref{dimensionless_Kernel_micro}, we conclude that, 
\beq
\sum_i {\cal R}^{\mu\nu}_{ij}(t-t') = 0 \ ,
\label{calippazzo}
\eeq
which implies that the second and third term in \eqref{eq:mag} are zero.
For what concerns the first term in \eqref{eq:mag}, we notice that it} can be rewritten as,
\begin{equation}
 \hbar^{-1} \sum_i \left (\dfrac{\partial 
H_{\mathrm{eff}}}{\partial \boldsymbol{\sigma}_i} \times \boldsymbol{\sigma}_i \right )=  \hbar^{-1} \sum_i \left (\dfrac{\partial 
H_{0}}{\partial \boldsymbol{\sigma}_i} \times \boldsymbol{\sigma}_i \right )- \hbar^{-1} \sum_{\boldsymbol{q}\beta} \dfrac{\alpha^2}{m \omega_{\boldsymbol{q}}^2}  f^{\beta}_{\boldsymbol{q}} \sum_i \left(  \dfrac{\partial   f^{\beta}_{-\boldsymbol{q}} }{\partial \boldsymbol{\sigma}_i} \times  \boldsymbol{\sigma}_i \right),
\end{equation}
where the first term in the r.h.s. gives 0, as we knew from the very beginning, while the second vanishes due to Eq. \eqref{eq_half_pseudo_laplacian}. The total magnetization is therefore strictly conserved during the dynamical evolution.

\subsection{Approximations for recovering DLT}
Up to this point, no approximation has been done, and both the conservation of the total magnetization and the FD relation are satisfied exactly. 
The equation of motion for the spin that we have obtained is
\begin{equation}\label{eq_of_motion_intermediate}
        \dfrac{d\sigma^{\mu}_i}{dt} = \hbar^{-1}\varepsilon_{\mu\nu\rho} \dfrac{\partial H_{\mathrm{eff}}}{\partial \sigma^{\nu}_i} \sigma^{\rho}_i -\hbar^{-1} \left(\dfrac{\alpha^2}{\hbar K}\right)\int_0^t dt' \sum_{j\nu}     2{\cal R}^{\mu \nu}_{ij}(t,t') \left( \dfrac{\partial H^{\mathrm{TOT}}}{\partial \sigma^{\nu}_j} \right)_{t'} +\xi^{\mu}_i \ .
\end{equation}
Eq. \eqref{eq_of_motion_intermediate} looks similar in structure to the equation of motion introduced in the main text in the DLT scheme: there is a Hamiltonian precession term, a conservative noise term, and a conservative relaxational term linked to the derivative of the Hamiltonian. However, there are several differences. First of all, the derivative of the Hamiltonian in the relaxational term involves the total Hamiltonian, which includes the phononic degrees of freedom making the equation itself not closed in the spin variables.  The precession term involves the effective Hamiltonian \eqref{def_H_eff}, instead of a simple spin exchange interaction. The relaxation term is proportional to something which, while having a zero mode, is not the discrete Laplacian. Finally, the memory kernel is not proportional to a Dirac delta distribution and the equation is therefore not Markovian. We address here the first three issues, while leaving the fourth to a separate section at the end of the SI.

\subsubsection{Approximation for $H^{\mathrm{TOT}}$ in the relaxational term}

Eq.\eqref{eq_of_motion_intermediate} involves the derivative of the total Hamiltonian with respect to the spin variables $\boldsymbol{\sigma}_j$. In the simpler case of a colloidal particle coupled to a bath of harmonic oscillators (see \cite{Zwanzig}), this term does not depend on the degrees of freedom of the thermal bath, leaving a closed equation for the colloidal particle. In the case of spin systems, though, this derivative includes the phononic degrees of freedom, so we have to make an approximation to close the equation. From the expression  of the total Hamiltonian  \eqref{H_tot} and the definition of $H_{\mathrm{eff}}$ we get 
\begin{equation}\label{approx_H_tot}
    \dfrac{\partial H^{\mathrm{TOT}}}{\partial \boldsymbol{\sigma}_j} = \dfrac{\partial H_{\mathrm{eff}}(\{{\boldsymbol \sigma}\} }{\partial \boldsymbol{\sigma}_j} - \alpha\sum_{\beta \boldsymbol{p}}   \dfrac{\partial f^{\beta}_{\boldsymbol{p}}}{\partial \boldsymbol{\sigma}_j} \cdot \left( Q^{\beta}_{-\boldsymbol{p}} -\dfrac{\alpha}{m\omega^2_{\boldsymbol{p}}}  f^{\beta}_{-\boldsymbol{p}} \right) \ .
\end{equation}
The phononic variables $Q^{\beta}_{-\boldsymbol{p}}(t)$ on the r.h.s. are random variables, since they depend on the initial condition $Q^{\beta}_{-\boldsymbol{p}}(0)$, which we have drawn from the canonical distribution.  If we assume that their dynamics is faster than the one of the spin variables we can approximate their time dependent value with the expected value with respect to their marginal equilibrium distribution at fixed $\boldsymbol{\sigma}$. Hence, the second term of Eq. \eqref{approx_H_tot} can be approximated with its average. Since the initial canonical distribution is invariant under the microcanonical dynamics, this average value is zero (see Eq. \eqref{average_Q}), thus giving
\textcolor{black}{
\beq
\dfrac{\partial H^{\mathrm{TOT}}}{\partial \boldsymbol{\sigma}_j} \simeq \dfrac{\partial H_{\mathrm{eff}}}{\partial \boldsymbol{\sigma}_j}
\eeq
}

\subsubsection{The effective Hamiltonian approximation}
The effective Hamiltonian $H_{\mathrm{eff}}$ appears now in both the precession and the relaxational term, and the equation of motion becomes,
\begin{equation}\label{eq_of_motion_intermediate_2}
        \dfrac{d\sigma^{\mu}_i}{dt} =  \hbar^{-1}\varepsilon_{\mu\nu\rho} \dfrac{\partial H_{\mathrm{eff}}}{\partial \sigma^{\nu}_i}  \sigma^{\rho}_i - \hbar^{-1} \left(\dfrac{\alpha^2}{\hbar K}\right)\int_0^t dt' \sum_{j\nu}     2{\cal R}^{\mu\nu}_{ij}(t,t') \left( \dfrac{\partial H_{\mathrm{eff}}}{\partial \sigma^{\nu}_j} \right)_{t'} +\xi^{\mu}_i .
\end{equation}
Eq. \eqref{eq_of_motion_intermediate_2} involves a Hamiltonian, which is a function of the spins only, and it is therefore self-consistent.
In the main text, though, we used an equation for the spin dynamics where only $H_0$ appears, in place of $H_{\mathrm{eff}}$.  In general, corrections to $H_0$ might be potentially relevant for the equilibrium probability distribution of the slower variables. However, this is not likely what happens in our case. Indeed, if we consider the difference between $H_0$ and  $H_{\mathrm{eff}}$ we have that,
\begin{equation}
    \Delta H= H_{\mathrm{eff}}- H_0 = -  \sum_{\boldsymbol{q}}  \frac{\alpha^2}{2m\omega_{\boldsymbol{q}}^2}  \boldsymbol{f}_{\boldsymbol{q}} \cdot  \boldsymbol{f}_{-\boldsymbol{q}}(\{\boldsymbol{\sigma}\}) , 
\end{equation}
and using, Eq. \eqref{eq_f},
\begin{equation}
    \Delta H= - \frac{\alpha^2}{2K}\sum_{\boldsymbol{q}} \sum_{ijkh}\dfrac{e^{i\boldsymbol{q}\cdot(\boldsymbol{r}_i-\boldsymbol{r}_k)}}{2\lambda_{\boldsymbol{q}}N} n_{ij} n_{kh} \hat{\boldsymbol{e}}_{ij}\cdot\hat{\boldsymbol{e}}_{kh} \left(\boldsymbol{\sigma}_i\cdot \boldsymbol{\sigma}_j \right)\left(\boldsymbol{\sigma}_k\cdot \boldsymbol{\sigma}_h \right). 
\end{equation}
The above expression is quite complicated, but we can see that both the  ground state of the ferromagnet (all the spins aligned) and the ground state of the antiferromagnet (all spins anti-aligned with their neighbors) yield $\Delta H = 0$ . Hence, we argue that the correction does not change any fundamental feature of the system and it is safe to neglect this contribution. This is also consistent with the fact that in the starting Hamiltonian we neglected all the 4-spins (or more) contributions keeping only 2-spins interactions. \textcolor{black}{With this approximation, we hence have, 
\begin{equation}
     H_{\mathrm{eff}} \simeq H_0 . 
\end{equation}
}

\subsubsection{Introducing the Laplacian}

We now focus on the site dependence of the memory kernel. As discussed in the previous sections, the kernel $exactly$ has a zero mode, which ensures the conservation law for the global magnetization. Adopting some additional approximations, we can show that it is proportional to the discrete Laplacian, with the appropriate $\boldsymbol{q}$-dependent coefficient. Let us call,
\begin{equation}
    \overline{\Lambda}_{ij}^{\mu\nu}(t,t')=\sum_{\beta} \left( \dfrac{\partial f^{\beta}_{-\boldsymbol{q}}}{\partial\boldsymbol{\sigma}_i} \times \boldsymbol{\sigma}_i \right)^{\mu}_t\left( \dfrac{\partial f^{\beta}_{\boldsymbol{q}}}{\partial\boldsymbol{\sigma}_j} \times \boldsymbol{\sigma}_j \right)^{\nu}_{t'},
\end{equation}
the site-dependent part of the memory kernel.  Using Eq.~\eqref{eq_half_pseudo_laplacian} we find, 
\begin{equation}\label{matrix_kernel}
\begin{aligned}
  \overline{\Lambda}_{ij}^{\mu\nu}(t,t') =\dfrac{1}{N} \sum_{\beta h k} \left(e^{i\boldsymbol{q} \cdot \boldsymbol{r}_i} -  e^{i\boldsymbol{q} \cdot \boldsymbol{r}_h} \right)\left(e^{i\boldsymbol{q} \cdot \boldsymbol{r}_j} -  e^{i\boldsymbol{q} \cdot \boldsymbol{r}_k} \right)  n_{ih} \hat{\boldsymbol{e}}_{ih}^{\beta}\ n_{jk} \ \hat{\boldsymbol{e}}_{jk}^{\beta} \ ( \boldsymbol{\sigma}_h \times \boldsymbol{\sigma}_i)_t^{\mu} ( \boldsymbol{\sigma}_k \times \boldsymbol{\sigma}_j)_{t'}^{\nu} .
\end{aligned}
\end{equation}
It is also convenient to introduce the following auxiliary quantity,
\begin{equation}
    \mathbf{L}_{hikj}^{\mu\nu}(t,t')= ( \boldsymbol{\sigma}_h \times \boldsymbol{\sigma}_i)_t^{\mu} ( \boldsymbol{\sigma}_k \times \boldsymbol{\sigma}_j)_{t'}^{\nu} .
\end{equation}
 First, we assume isotropy with respect to the spin vectorial components, so that this matrix becomes proportional to the identity in that space, 
\begin{equation}
    \mathbf{L}_{hikj}^{\mu\nu}(t,t')\simeq  \dfrac{1}{d} ( \boldsymbol{\sigma}_h \times \boldsymbol{\sigma}_i)_t \cdot ( \boldsymbol{\sigma}_k \times \boldsymbol{\sigma}_j)_{t'} \delta_{\mu\nu} \ . 
\end{equation}
 By applying the vectorial identity $ (\boldsymbol{a}\times\boldsymbol{b}) \cdot ( \boldsymbol{c} \times \boldsymbol{d} ) = (\boldsymbol{a} \cdot \boldsymbol{c}) (\boldsymbol{b} \cdot \boldsymbol{d} ) - (\boldsymbol{a} \cdot \boldsymbol{d} ) (\boldsymbol{b} \cdot \boldsymbol{c})$
we can rewrite this quantity in a more useful form,
\begin{equation}
\begin{aligned}
    \mathbf{L}_{hikj}^{\mu\nu}(t,t') \simeq \dfrac{1}{d} \left[ \left(\boldsymbol{\sigma}_h(t) \cdot \boldsymbol{\sigma}_k(t') \right) \left(\boldsymbol{\sigma}_i(t) \cdot \boldsymbol{\sigma}_j(t') \right) - \left(\boldsymbol{\sigma}_h(t) \cdot \boldsymbol{\sigma}_j(t') \right) \left(\boldsymbol{\sigma}_i(t) \cdot \boldsymbol{\sigma}_k(t') \right)   \right] \delta_{\mu\nu} .
    \end{aligned}
\end{equation}
We then approximate the above expression with its expected value with respect to the stationary spin distribution,
\begin{equation}
    \begin{aligned}
        \mathbf{L}_{hikj}^{\mu\nu}(t,t') 
         \simeq \dfrac{1}{d} \left[ \langle\left(\boldsymbol{\sigma}_h(t) \cdot \boldsymbol{\sigma}_k(t') \right) \left(\boldsymbol{\sigma}_i(t) \cdot \boldsymbol{\sigma}_j(t') \right)\rangle - \langle\left(\boldsymbol{\sigma}_h(t) \cdot \boldsymbol{\sigma}_j(t') \right) \left(\boldsymbol{\sigma}_i(t) \cdot \boldsymbol{\sigma}_k(t') \right)\rangle   \right] \delta_{\mu\nu} ,
    \end{aligned}
\end{equation}
and we consider spin correlations to be negligible unless the scalar product is taken between two spins on the same site (i.e. for $i=j$ and $h=k$ in the first term and $h=j$ and $i=k$ in the second term),
\begin{equation}\label{ciao}
    \begin{aligned}
      \mathbf{L}_{hikj}^{\mu\nu}(t,t')   & \simeq \dfrac{1}{d} \left[ \langle\left(\boldsymbol{\sigma}_h(t) \cdot \boldsymbol{\sigma}_h(t') \right) \left(\boldsymbol{\sigma}_i(t) \cdot \boldsymbol{\sigma}_i(t') \right)\rangle \delta_{hk}\delta_{ij} - \langle\left(\boldsymbol{\sigma}_j(t) \cdot \boldsymbol{\sigma}_j(t') \right) \left(\boldsymbol{\sigma}_i(t) \cdot \boldsymbol{\sigma}_i(t') \right)\rangle \delta_{jh}\delta_{ik}   \right] \delta_{\mu\nu}\\
        & =  \dfrac{1}{d} \left[ \langle\left(\boldsymbol{\sigma}_h(t) \cdot \boldsymbol{\sigma}_h(t') \right) \left(\boldsymbol{\sigma}_i(t) \cdot \boldsymbol{\sigma}_i(t') \right)\rangle \left(\delta_{hk}\delta_{ij} -\delta_{jh}\delta_{ik} \right) \right] \delta_{\mu\nu} .
   \end{aligned}
\end{equation}
A four-point correlation function of the spins appears in Eq. \eqref{ciao}, namely,
   \begin{equation}
   C^{(4)}(t-t')=\langle\left(\boldsymbol{\sigma}_h(t) \cdot \boldsymbol{\sigma}_h(t') \right) \left(\boldsymbol{\sigma}_i(t) \cdot \boldsymbol{\sigma}_i(t') \right)\rangle
   \end{equation} 
 where no site dependence is present in the l.h.s because $i$ and $h$ are always first neighbours (see \eqref {matrix_kernel}) and the resulting expected value is therefore the same for all the first neighbours pairs.  The time dependences appear only as a time difference, because expected values are taken with respect to the stationary distribution. The actual shape of this function is not important: we only assume that the time dependence is much slower than the phononic dynamics, so that we can assume it to be constant in time and approximate it with its value in 0, namely $C^{(4)}(t-t')\simeq C^{(4)}(0)=1$.
Now we can use the above approximations in formula \eqref{matrix_kernel} and obtain,
\begin{equation}
\begin{aligned}
  \overline{\Lambda}_{ij}^{\mu\nu}(t,t')    \simeq \overline{\Lambda}_{ij}^{\mu\nu}    \simeq & \dfrac{1}{Nd} \sum_{\beta h k} \left(e^{i\boldsymbol{q} \cdot \boldsymbol{r}_i} -  e^{i\boldsymbol{q} \cdot \boldsymbol{r}_h} \right)\left(e^{i\boldsymbol{q} \cdot \boldsymbol{r}_j} -  e^{i\boldsymbol{q} \cdot \boldsymbol{r}_k} \right)  n_{ih} \ \hat{\boldsymbol{e}}_{ih}^{\beta}\ n_{jk} \ \hat{\boldsymbol{e}}_{jk}^{\beta} \ \delta_{\mu\nu} \left(\delta_{hk}\delta_{ij} -\delta_{jh}\delta_{ik} \right) =\\
   =& \dfrac{1}{Nd} \Bigg[ \sum_{\beta h} \left(e^{i\boldsymbol{q} \cdot \boldsymbol{r}_i} -  e^{i\boldsymbol{q} \cdot \boldsymbol{r}_h} \right)\left(e^{i\boldsymbol{q} \cdot \boldsymbol{r}_i} -  e^{i\boldsymbol{q} \cdot \boldsymbol{r}_h} \right)  n_{ih} \ \hat{\boldsymbol{e}}_{ih}^{\beta}\ \hat{\boldsymbol{e}}_{ih}^{\beta} \ \delta_{ij}    \\
    - & \sum_{\beta h} \left(e^{i\boldsymbol{q} \cdot \boldsymbol{r}_i}- e^{i\boldsymbol{q} \cdot \boldsymbol{r}_h} \right)\left(e^{i\boldsymbol{q} \cdot \boldsymbol{r}_h} -  e^{i\boldsymbol{q} \cdot \boldsymbol{r}_i} \right)  n_{ih} \ \hat{\boldsymbol{e}}_{ih}^{\beta}\  \hat{\boldsymbol{e}}_{hi}^{\beta}\  \delta_{jh} \Bigg] \delta_{\mu\nu} ,\\
   \end{aligned}
   \end{equation}
   We remind that $\hat{\boldsymbol{e}}_{hi}=-\hat{\boldsymbol{e}}_{ih}$, thus finding,
   \begin{equation}
   \begin{aligned}
   \overline{\Lambda}_{ij}^{\mu\nu} \simeq &  \dfrac{1}{Nd} \Bigg[ \sum_{h} |e^{i\boldsymbol{q} \cdot \boldsymbol{r}_i} -  e^{i\boldsymbol{q} \cdot \boldsymbol{r}_h} |^2 n_{ih} \delta_{ij} - |e^{i\boldsymbol{q} \cdot \boldsymbol{r}_i}- e^{i\boldsymbol{q} \cdot \boldsymbol{r}_h}|^2  n_{ih}\delta_{jh} \Bigg] \delta_{\mu\nu} =\\
   = & \dfrac{1}{Nd} \Bigg[ \sum_{h} 2(1-\cos(\boldsymbol{q} \cdot (\boldsymbol{r}_i-\boldsymbol{r}_h)) n_{ih} \delta_{ij} - 2(1-cos(\boldsymbol{q} \cdot (\boldsymbol{r}_i-\boldsymbol{r}_j))  n_{ij} \Bigg] \delta_{\mu\nu} =\\
   = & \dfrac{2}{Nd} \Bigg[ \lambda_{\boldsymbol{q}} \delta_{ij} - (1-\cos(\boldsymbol{q} \cdot (\boldsymbol{r}_i-\boldsymbol{r}_j))  n_{ij} \Bigg] \delta_{\mu\nu} \ .
\end{aligned}
\end{equation}
The indexes $i$ and $j$ identify a well defined direction $\alpha=\alpha(i,j)$, so the last equation reads,
\begin{equation}
   \overline{\Lambda}_{ij}^{\mu\nu} = \dfrac{2}{Nd} \Bigg[ \lambda_{\boldsymbol{q}} \delta_{ij} - (1-\cos(q^{\alpha(i,j)} l))  n_{ij} \Bigg] \delta_{\mu\nu} \ .
\end{equation}
Lastly, we make the assumption of isotropy in the $\boldsymbol{q}$-space,
    \begin{equation}
        (1-\cos(q^{\alpha(i,j)} l)) \simeq \dfrac{1}{d} \sum_{\beta=1}^d (1-\cos(q^{\beta} l))\equiv \dfrac{1}{n_c}\lambda_{\boldsymbol{q}}  \hspace{1cm} \forall \alpha ,
    \end{equation}
hence we end up with,
\begin{equation}
\begin{aligned}
   \overline{\Lambda}_{ij}^{\mu\nu} & \simeq \dfrac{2}{Nd n_c}  \lambda_{\boldsymbol{q}} \left( n_c \delta_{ij} -  n_{ij} \right) \delta_{\mu\nu}  =\\
    & = \dfrac{1}{Nd^2}  \lambda_{\boldsymbol{q}} \Lambda_{ij} \delta_{\mu\nu} \ .
    \end{aligned}
\end{equation}
With this expression, we finally recover the Laplacian matrix appearing in the DLT scheme. \textcolor{black}{We now can reduce the memory kernel introduced before to,
\begin{equation}\label{kernel_reduced}
    {\cal R}^{\mu\nu}_{ij}(t-t') \simeq \Lambda_{ij}\delta_{\mu\nu}{\cal R}(t-t') ,
\end{equation}
}
where,
\begin{equation}\label{kernel_micro_scalar}
    {\cal R}(t-t') = \sum_{\boldsymbol{q}} \frac{1}{4N d^2} \cos(\omega_{\boldsymbol{q}}(t-t')) ,
\end{equation}
  And we can rewrite, after all the approximations discussed so far, Eq.\eqref{eq_of_motion_intermediate_2} as,
\begin{equation}\label{eq_of_motion_intermediate_3}
        \dfrac{d \boldsymbol{\sigma}_i}{dt} = \hbar^{-1} \dfrac{\partial H_{0}}{\partial \boldsymbol{\sigma}_i} \times \boldsymbol{\sigma}_i - \hbar^{-1} \left(\dfrac{\alpha^2}{\hbar K}\right) \sum_{j}     \Lambda_{ij}  \int_0^t dt' 2{\cal R}(t-t')\left( \dfrac{\partial H_{0}}{\partial \boldsymbol{\sigma}_j} \right)_{t'} +\boldsymbol{\xi}_i ,
\end{equation}
with the noise correlation, 
\begin{equation}\label{eq:correlation_noise_kernel_bisanzio}
    \langle \xi^{\mu}_i(t) \xi^{\nu}_j(t') \rangle = 2k_BT  \hbar^{-1} \left(\dfrac{\alpha^2}{\hbar K}\right) \Lambda_{ij} \delta_{\mu\nu} {\cal R}(t-t').
\end{equation}

This dynamics is now nearly identical to DLT. The only remaining difference is that the dimensionless memory kernel ${\cal R}(t-t')$ is not Markovian, as we hinted in our shortened derivation in the main text using dimensional analysis. The time dependence of the kernel is in general non-trivial and it depends on the details of the phononic spectrum, and on the phonon dynamics. So far, we have considered the case where the system is isolated and the dynamics is fully microcanonic. In the next section, we compute the time dependence of the memory kernel when the phononic degrees of freedom (but not the spins) are coupled with an external thermal bath.

\subsection{Phonons coupled to an external thermal bath}\label{sec:canonical} 
     
\subsubsection{Equations of motion}
In this Section we perform the same computation as in Sec. \ref{sec:eq_motion_micro}, but considering this time the vibrational degrees of freedom coupled to an external thermal bath.  The spin variables, on the other hand, still follow Hamiltonian equations of motion and feel the presence of the thermal bath only via their interaction with the lattice. We use for the phonons a standard Gaussian white thermal bath, with friction coefficient $\eta$ and temperature $T$. Instead of Eq. \eqref{eq_of_motion_phonons_micro} we therefore have,

\begin{equation}
    m \dfrac{d^2\boldsymbol{Q}_{\boldsymbol{q}}}{dt^2} + \eta \dfrac{d\boldsymbol{Q}_{\boldsymbol{q}}}{dt} + 2K \lambda_{\boldsymbol{q}} \boldsymbol{Q}_{\boldsymbol{q}} = \boldsymbol{\zeta}_{\boldsymbol{q}} + \alpha\boldsymbol{f}_{\boldsymbol{q}},
\end{equation}
where the bath noise satisfies,
\begin{equation}
    \langle\boldsymbol{\zeta}_{\boldsymbol{q}}\rangle= 0,
\end{equation}
\begin{equation}
    \langle\boldsymbol{\zeta}_{\boldsymbol{q}}(t)\boldsymbol{\zeta}_{\boldsymbol{p}}(t')\rangle= 2\eta T \delta_{q,-p}\delta(t-t') .
\end{equation}
 There are three relevant (inverse) timescales in this equation, i.e.
 \begin{equation}
     \omega_{\boldsymbol{q}}= \sqrt{\dfrac{2K}{m}\lambda_{\boldsymbol{q}}} ,
\quad\quad\quad\quad
     \gamma=\dfrac{\eta}{2m} ,
\quad \quad\quad\quad
     \overline{\omega}_{\boldsymbol{q}}=\sqrt{\omega^2_{\boldsymbol{q}}-\gamma^2} .
 \end{equation}
The solution of the homogeneous equation is not relevant, since it only gives an exponentially decaying transient, and for large enough times that system forgets the initial conditions. Setting the initial condition at $t_0=-\infty$ the solution is therefore given by,
\begin{equation}
\begin{aligned}
\boldsymbol{Q}_{\boldsymbol{q}}(t) = & \dfrac{\alpha}{m \overline{\omega}_{\boldsymbol{q}}} \int_{-\infty}^t dt' e^{-\gamma(t-t')}\sin(\overline{\omega}_{\boldsymbol{q}}(t-t')) \boldsymbol{f}_{\boldsymbol{q}}(t')+  \dfrac{1}{m \overline{\omega}_{\boldsymbol{q}}} \int_{-\infty}^t dt' e^{-\gamma(t-t')}\sin(\overline{\omega}_{\boldsymbol{q}}(t-t')) \boldsymbol{\zeta}_{\boldsymbol{q}}(t') \\
= & \dfrac{\alpha}{m\omega_{\boldsymbol{q}}^2}\boldsymbol{f}_{\boldsymbol{q}}(t) - \dfrac{\alpha}{m \omega_{\boldsymbol{q}}^2} \int_{-\infty}^t dt' e^{-\gamma(t-t')} \left(\cos(\overline{\omega}_{\boldsymbol{q}}(t-t')) - \dfrac{\gamma}{\overline{\omega}_{\boldsymbol{q}}    }\sin(\overline{\omega}_{\boldsymbol{q}}(t-t'))  \right) \dfrac{d}{dt'}\boldsymbol{f}_{\boldsymbol{q}}(t')\\
+ & \dfrac{1}{m  \overline{\omega}_{\boldsymbol{q}}} \int_{-\infty}^t dt' e^{-\gamma(t-t')}\sin(\overline{\omega}_{\boldsymbol{q}}(t-t')) \boldsymbol{\zeta}_{\boldsymbol{q}}(t') .
\end{aligned}
\end{equation}
The above equations can then be inserted back in the equations of motion for the $\boldsymbol{\sigma}$ variables, 
\begin{equation}
\begin{aligned}
 \dfrac{d \boldsymbol{\sigma}_i}{dt} = &  \hbar^{-1}\dfrac{\partial H}{\partial \boldsymbol{\sigma}_i} \times \boldsymbol{\sigma}_i=\hbar^{-1} \dfrac{\partial H_0}{\partial \boldsymbol{\sigma}_i} \times \boldsymbol{\sigma}_i - \dfrac{\alpha}{\hbar} \sum_{\boldsymbol{q}\beta}  Q^{\beta}_{\boldsymbol{q}}  \dfrac{\partial   f^{\beta}_{-\boldsymbol{q}}(\{\boldsymbol{\sigma} \}) }{\partial \boldsymbol{\sigma}_i}  \times \boldsymbol{\sigma}_i \ . \\
\end{aligned}
\end{equation}
Proceeding as in Sec. \ref{sec:eq_motion_micro} we find, 
\begin{equation}
    \dfrac{d \boldsymbol{\sigma}_i}{dt} = \hbar^{-1} \dfrac{\partial H_{\mathrm{eff}}}{\partial \boldsymbol{\sigma}_i} \times \boldsymbol{\sigma}_i - \boldsymbol{\Xi}_i\left[\{\boldsymbol{\sigma}\}\right] +\boldsymbol{\xi}_i ,
\end{equation}
where $H_{\mathrm{eff}}$ is the same defined in Eq. \eqref{def_H_eff},  but the relaxational and the noise terms now have a different expression,
\begin{equation}\label{def_relaxational_canonical}
    \boldsymbol{\Xi}_i\left[\{\boldsymbol{\sigma}\}\right] = - \sum_{\boldsymbol{q}\beta}   \dfrac{\alpha^2}{m \hbar \omega_{\boldsymbol{q}}^2} \int_{-\infty}^t dt' e^{-\gamma(t-t')} \left(\cos(\overline{\omega}_{\boldsymbol{q}}(t-t')) - \dfrac{\gamma}{\overline{\omega}_{\boldsymbol{q}}    }\sin(\overline{\omega}_{\boldsymbol{q}}(t-t'))  \right)\dfrac{d}{dt'}f^{\beta}_{\boldsymbol{q}}(t') \left(\dfrac{\partial   f^{\beta}_{-\boldsymbol{q}}(\{\boldsymbol{\sigma} \}) }{\partial \boldsymbol{\sigma}_i} \times \boldsymbol{\sigma}_i \right)_t,
\end{equation}
 and,
\begin{equation}\label{def_noise_canonical}
\boldsymbol{\xi}_i = \dfrac{\alpha}{\hbar} \sum_{\boldsymbol{q} \beta} \left( \dfrac{1}{m  \overline{\omega}_{\boldsymbol{q}}} \int_{-\infty}^t dt' e^{-\gamma(t-t')}\sin(\overline{\omega}_{\boldsymbol{q}}(t-t')) \zeta^{\beta}_{\boldsymbol{q}}(t') \right) \bigg(  \dfrac{\partial f^{\beta}_{-\boldsymbol{q}}(\{\boldsymbol{\sigma} \})}{\partial \boldsymbol{\sigma}_i} \times \boldsymbol{\sigma}_i \bigg)_t .
\\
\end{equation}

\subsubsection{Noise}

Let us start by considering the noise term $\boldsymbol{\xi}_i$ defined in \eqref{def_noise_canonical}. In the microcanonical treatment of Sec. \ref{sec:eq_motion_micro}, the noise was a fluctuating variable due to the presence of the initial conditions, which were drawn randomly with a canonical distribution. Here, on the other hand, the initial conditions are not relevant and the fluctuating nature of this term is due to the presence of the stochastic variable $\zeta^{\beta}_{\boldsymbol{q}}$ representing the external thermal bath coupled to phonons. 
Taking the average of Eq.\eqref{def_noise_canonical} over the thermal bath at fixed $\boldsymbol{\sigma}$ we get,
\begin{equation}\label{eq:average_noise_zero}
    \langle \boldsymbol{\xi}_i \rangle = 0,
\end{equation}
and, for the noise correlator,
\begin{equation}
\begin{aligned}
     \langle \xi^{\mu}_i(t) \xi^{\nu}_j(t') \rangle & = \dfrac{\alpha^2}{\hbar^2} \sum_{\alpha\beta\boldsymbol{q}\boldsymbol{q}'} \langle \bigg[ \bigg( \dfrac{1}{m  \overline{\omega}_{\boldsymbol{q}}} \int_{-\infty}^t ds e^{-\gamma(t-s)}\sin(\overline{\omega}_{\boldsymbol{q}}(t-s)) \zeta^{\alpha}_{\boldsymbol{q}}(s) \bigg)\\
     &\bigg( \dfrac{1}{m\overline{\omega}_{\boldsymbol{q}'}}\int_{-\infty}^{t'} ds' e^{-\gamma(t'-s')}\sin(\overline{\omega}_{\boldsymbol{q}'}(t'-s')) \zeta^{\beta}_{\boldsymbol{q}'}(s') \bigg) \bigg] \rangle \bigg(  \dfrac{\partial f^{\alpha}_{-\boldsymbol{q}}(\{\boldsymbol{\sigma} \})}{\partial \boldsymbol{\sigma}_i} \times \boldsymbol{\sigma}_i \bigg)^{\mu}_{t} \bigg(  \dfrac{\partial f^{\beta}_{-\boldsymbol{q}'}(\{\boldsymbol{\sigma} \}) }{\partial \boldsymbol{\sigma}_j} \times  \boldsymbol{\sigma}_j \bigg)^{\nu}_{t'}\\
    &= \sum_{\beta \boldsymbol{q}} \dfrac{4\gamma\alpha^2 k_B T}{m \hbar^2 \overline{\omega}^2_{\boldsymbol{q}}} \bigg[\int_{-\infty}^{\mathrm{min}(t,t')} ds e^{-\gamma(t+t'-2s)}\sin(\overline{\omega}_{\boldsymbol{q}}(t-s))\sin(\overline{\omega}_{\boldsymbol{q}}(t'-s))   \bigg] \\
    & \bigg(  \dfrac{\partial f^{\beta}_{-\boldsymbol{q}}(\{\boldsymbol{\sigma} \})}{\partial \boldsymbol{\sigma}_i} \times \boldsymbol{\sigma}_i \bigg)^{\mu}_t \bigg(  \dfrac{\partial f^{\beta}_{\boldsymbol{q}}(\{\boldsymbol{\sigma} \}) }{\partial \boldsymbol{\sigma}_j} \times  \boldsymbol{\sigma}_j \bigg)^{\nu}_{t'} =\\ 
     = & \sum_{\beta \boldsymbol{q}} \dfrac{4\gamma\alpha^2 k_B T}{m \hbar^2 \overline{\omega}^2_{\boldsymbol{q}}} \bigg[e^{-\gamma(t+t')}\int_{-\infty}^{\mathrm{min}(t,t')} ds e^{2\gamma s}\dfrac{1}{2}\bigg( \cos(\overline{\omega}_{\boldsymbol{q}}(t-t')) - \cos(\overline{\omega}_{\boldsymbol{q}}(t+t'-2s)) \bigg)  \bigg]\\
     & \bigg(  \dfrac{\partial f^{\beta}_{-\boldsymbol{q}}(\{\boldsymbol{\sigma} \})}{\partial \boldsymbol{\sigma}_i} \times \boldsymbol{\sigma}_i \bigg)^{\mu}_t \bigg(  \dfrac{\partial f^{\beta}_{\boldsymbol{q}}(\{\boldsymbol{\sigma} \}) }{\partial \boldsymbol{\sigma}_j} \times  \boldsymbol{\sigma}_j \bigg)^{\nu}_{t'} =\\
     = & \sum_{\beta \boldsymbol{q}} \dfrac{2\alpha^2 k_B T}{m\hbar^2\overline{\omega}^2_{\boldsymbol{q}}} \frac{1}{2}\bigg[e^{-\gamma|t-t'|}\cos(\overline{\omega}_{\boldsymbol{q}}(t-t')) -\dfrac{\gamma}{{\omega}^2_{\boldsymbol{q}}}  e^{-\gamma|t-t'|}\left(  \gamma\cos(\overline{\omega}_{\boldsymbol{q}}(t-t'))   +  \overline{\omega}_{\boldsymbol{q}} \sin(\overline{\omega}_{\boldsymbol{q}}|t-t'|) \right) \bigg]\\
     & \bigg(  \dfrac{\partial f^{\beta}_{-\boldsymbol{q}}(\{\boldsymbol{\sigma} \})}{\partial \boldsymbol{\sigma}_i} \times \boldsymbol{\sigma}_i \bigg)^{\mu}_t \bigg(  \dfrac{\partial f^{\beta}_{\boldsymbol{q}}(\{\boldsymbol{\sigma} \}) }{\partial \boldsymbol{\sigma}_j} \times  \boldsymbol{\sigma}_j \bigg)^{\nu}_{t'} =\\
    =  \sum_{\beta \boldsymbol{q}} & \dfrac{\alpha^2 k_B T}{m \hbar^2 \omega^2_{\boldsymbol{q}}} e^{-\gamma|t-t'|}\bigg(\cos(\overline{\omega}_{\boldsymbol{q}}|t-t'|) -\dfrac{\gamma}{\overline{\omega}_{\boldsymbol{q}}} \sin(\overline{\omega}_{\boldsymbol{q}}|t-t'|) \bigg) \bigg(  \dfrac{\partial f^{\beta}_{-\boldsymbol{q}}(\{\boldsymbol{\sigma} \})}{\partial \boldsymbol{\sigma}_i} \times \boldsymbol{\sigma}_i \bigg)^{\mu}_t \bigg(  \dfrac{\partial f^{\beta}_{\boldsymbol{q}}(\{\boldsymbol{\sigma} \}) }{\partial \boldsymbol{\sigma}_j} \times  \boldsymbol{\sigma}_j \bigg)^{\nu}_{t'} .
\end{aligned}
\end{equation}
Hence, we have again that the $\{\boldsymbol{\xi}_i\}$ are Gaussian random variables with 0 mean and non trivial correlations in space and time. We can in this case identify the dimensionless kernel,
\begin{equation}\label{dimensionless_kernel_canonical}
    {\cal R}^{\mu\nu}_{ij}(t,t') =\sum_{\beta \boldsymbol{q}} \dfrac{1}{ 4\lambda_{\boldsymbol{q}}} e^{-\gamma|t-t'|}\bigg(\cos(\overline{\omega}_{\boldsymbol{q}}|t-t'|) -\dfrac{\gamma}{\overline{\omega}_{\boldsymbol{q}}} \sin(\overline{\omega}_{\boldsymbol{q}}|t-t'|) \bigg)\\
      \bigg(  \dfrac{\partial f^{\beta}_{-\boldsymbol{q}}(\{\boldsymbol{\sigma} \})}{\partial \boldsymbol{\sigma}_i} \times \boldsymbol{\sigma}_i \bigg)^{\mu}_t \bigg(  \dfrac{\partial f^{\beta}_{\boldsymbol{q}}(\{\boldsymbol{\sigma} \}) }{\partial \boldsymbol{\sigma}_j} \times  \boldsymbol{\sigma}_j \bigg)^{\nu}_{t'} ,
\end{equation}
and rewrite the correlator as,
\begin{equation}\label{eq:correlation_noise_kernel_bis}
    \langle \xi^{\mu}_i(t) \xi^{\nu}_j(t') \rangle =  2k_BT \hbar^{-1} \left(\dfrac{\alpha^2}{\hbar K}\right) {\cal R}^{\mu\nu}_{ij}(t,t') .
\end{equation}
This last result generalizes the one of Eq. \eqref{dimensionless_Kernel_micro}, while the microcanonical result is recovered when $\gamma=0$.
          
\subsubsection{Relaxational term and the FD relation}
Let us now consider the relaxational term $\boldsymbol{\Xi}_i\left[\{\boldsymbol{\sigma}\}\right]$ in Eq. \eqref{def_relaxational_canonical}.
Since $t>t'$ inside the integral, we can put a modulus in the argument of the exponential, of the cosine and of the sine. Then, again, we can use Eq. \eqref{eq:time_derivative_f} to expand the time derivative of $f^{\beta}_{\boldsymbol{q}}$, and we find,
\begin{equation}
\begin{aligned}
\boldsymbol{\Xi}_i\left[\{\boldsymbol{\sigma}\}\right] = & \sum_{j \boldsymbol{q}, \beta}   \dfrac{\alpha^2}{m\hbar \omega_{\boldsymbol{q}}^2} \int_{-\infty}^t dt' e^{-\gamma|t-t'|} \left(\cos(\overline{\omega}_{\boldsymbol{q}}|t-t'|) - \dfrac{\gamma}{\overline{\omega}_{\boldsymbol{q}}    }\sin(\overline{\omega}_{\boldsymbol{q}}|t-t'|)  \right)\\
& \left( \dfrac{\partial   f^{\beta}_{-\boldsymbol{q}}(\{\boldsymbol{\sigma} \}) }{\partial \boldsymbol{\sigma}_i} \times \boldsymbol{\sigma}_i \right)_t \left( \dfrac{\partial f^{\beta}_{\boldsymbol{q}}}{\partial\boldsymbol{\sigma}_j} \times \boldsymbol{\sigma}_j \right)_{t'} \cdot \left( \hbar^{-1} \dfrac{\partial H^{\mathrm{TOT}}}{\partial \boldsymbol{\sigma}_j} \right) ,
\end{aligned}
\end{equation}
or, equivalently, using the definition of ${\cal R}^{\mu\nu}_{ij}(t,t')$ given in \eqref{dimensionless_kernel_canonical},
\begin{equation}\label{relaxational_kernel_canonical_indices}
    \Xi_i^{\mu}= \hbar^{-1} \left(\dfrac{\alpha^2}{\hbar K}\right)\int_{-\infty}^t dt' \sum_{j,\nu}   2  {\cal R}^{\mu\nu}_{ij}(t,t') \left( \dfrac{\partial H^{\mathrm{TOT}}}{\partial \sigma^{\nu}_j} \right)_{t'} ,
\end{equation}
 Also in this case therefore the memory function of the relaxational term is the same as the correlation of the  noise, up to a factor $k_BT$. Hence, the FD relation is satisfied also with this dynamics, which allows recovering the equilibrium behavior at temperature $T$. The conservation law of the total magnetization is also ensured with this thermal bath, as the the site-dependent part of the kernel defined in \eqref{dimensionless_kernel_canonical} is the same as in \eqref{eq:noise_kernel_micro}. 

Finally, we can perform all the approximations that we have discussed for the isolated system case. The final result of this computation is the same as the previous one, except for the time dependence of  ${\cal R}^{\mu\nu}_{ij}(t,t')$. Hence, we end up with,
\begin{equation}
\boldsymbol{\Xi}_i\left[\{\boldsymbol{\sigma}\}\right] =  \hbar^{-1}  \left(\dfrac{\alpha^2}{\hbar K}\right) \sum_{j}\Lambda_{ij}\int_{-\infty}^t dt'    2  {\cal R}(t-t') \left(\dfrac{\partial H^{\mathrm{TOT}}}{\partial \sigma^{\nu}_j} \right)_{t'} ,
\end{equation}
where now the scalar memory kernel is given by
\begin{equation}
\begin{aligned}
{\cal R}(t-t') = & \sum_{ \boldsymbol{q}}   \dfrac{1}{ 4Nd^2}  e^{-\gamma|t-t'|} \left(\cos(\overline{\omega}_{\boldsymbol{q}}|t-t'|) - \dfrac{\gamma}{\overline{\omega}_{\boldsymbol{q}}    }\sin(\overline{\omega}_{\boldsymbol{q}}|t-t'|)  \right) \ .\\
\end{aligned}
\end{equation}

\subsection{Summary of phonons marginalization}
Let us summarize the main results obtained so far. We started from the equations of motion for a system of spin degrees of freedom following a precession dynamics and interacting with the vibrational degrees of freedom of the lattice. We explicitly solved the equations for the phonons, as a function of the spins. This procedure lead, after some approximations, to the following effective equations of motion for the spins only\footnote
{We set the initial condition $t_0=-\infty$ and write the two cases, isolated system and phonons coupled to an external bath, in the same way.},
    \begin{equation}\label{eq_of_motion_intermediate_3_bis}
        \dfrac{d \boldsymbol{\sigma}_i}{dt} =\lbrace H_0, \boldsymbol{\sigma}_i \rbrace -\hbar^{-1} \left(\dfrac{\alpha^2}{\hbar K}\right)\sum_{j}     \Lambda_{ij}\int_{-\infty}^t dt' 2{\cal R}(t-t') \left( \dfrac{\partial H_{0}}{\partial \boldsymbol{\sigma}_j} \right)_{t'} +\boldsymbol{\xi}_i ,
\end{equation}
where  $\boldsymbol{\xi}_i$ is a Gaussian  noise, with zero mean and correlation given by,
\begin{equation}\label{eq:correlation_noise_kernel_tris}
    \langle \xi^{\mu}_i(t) \xi^{\nu}_j(t') \rangle =  2k_BT \hbar^{-1} \left(\dfrac{\alpha^2}{\hbar K}\right) \Lambda_{ij}\delta_{\mu\nu} {\cal R}(t-t') \ .
\end{equation}
${\cal R}(t-t')$ is a dimensionless non-Markovian memory kernel dependent on the details of the phonons. For an isolated system, we have,
\begin{equation}
\begin{aligned}
{\cal R}(t-t') = & \sum_{ \boldsymbol{q}}   \dfrac{1}{ 4Nd^2}   \cos(\omega_{\boldsymbol{q}}t-t') \ ,
\\
\end{aligned}
\end{equation}
while for a system where phonons (but not spins) are in contact with a thermal bath, we have,
\begin{equation}
\begin{aligned}
{\cal R}(t-t') = & \sum_{ \boldsymbol{q}}   \dfrac{1}{ 4Nd^2}  e^{-\gamma|t-t'|} \left(\cos(\overline{\omega}_{\boldsymbol{q}}|t-t'|) - \dfrac{\gamma}{\overline{\omega}_{\boldsymbol{q}}    }\sin(\overline{\omega}_{\boldsymbol{q}}|t-t'|)  \right).\\
\end{aligned}
\end{equation}

\subsection{Discrete-time Markovian approximation }

Because of the memory kernel in Eq.\eqref{eq_of_motion_intermediate_3_bis}, what we have obtained so far is a non-Markovian thermostat, while the DLT presented in the main text is Markovian. In this section we show  that, under appropriate conditions, we can recover a Markovian dynamics by discretizing the continuous stochastic process over a sufficiently long time scale. Every stochastic differential equation must ultimately be brought back to its discrete time version for it to make sense, so we rewrite \eqref{eq_of_motion_intermediate_3_bis} as,
    \begin{equation}\label{eq_of_motion_differential}
        d \boldsymbol{\sigma}_i =\lbrace H_0, \boldsymbol{\sigma}_i \rbrace dt - \left[ \hbar^{-1} \left(\dfrac{\alpha^2}{\hbar K}\right)\sum_{j}     \Lambda_{ij}\int_{-\infty}^t dt' 2{\cal R}(t-t') \left( \dfrac{\partial H_{0}}{\partial \boldsymbol{\sigma}_j} \right)_{t'}\right]dt  +d\boldsymbol{{w}}_i ,
\end{equation}
where $d\boldsymbol{{w}}_i={\boldsymbol{\xi}}_i dt $ is the noise on scale $dt$. The microscopic timescale $dt$ has to be interpreted as the minimal physical time increment over which the system's dynamics takes place. The derivation of the previous sections shows that on this minimal timescale the noise correlator has memory, i.e. $\langle d\boldsymbol{{w}}_i(t)\cdot d\boldsymbol{{w}}_i(t')\rangle\sim {\cal R}(t-t') dt^2$.
We can however ask whether, when considering the dynamics on scales larger than this minimal increment, a Markovian process can be recovered. Let us assume that the spins vary in time slowly compared with the time span of ${\cal R}(t-t')$. In this case, we immediately see that a simplification occurs in Eq. \eqref{eq_of_motion_differential}, because the quantity $\partial H_{0} / \partial {\boldsymbol{\sigma}_j}$ only depends on the spin variables and can be brought out of the integral.
%
If we then define,
\begin{equation}
    \tau_{\mathrm{m}} = \int_{-\infty}^{+\infty} ds {\cal R}(s)=\int_{-\infty}^0 ds 2{\cal R}(s),
\end{equation}
Eq. \eqref{eq_of_motion_differential} can be rewritten as,
    \begin{equation}\label{eq_of_motion_differential_3}
        d \boldsymbol{\sigma}_i =\lbrace H_0, \boldsymbol{\sigma}_i \rbrace dt - \left[ \hbar^{-1} \left(\dfrac{\alpha^2}{\hbar K}\right)\tau_{\mathrm{m}}\sum_{j}     \Lambda_{ij}\left( \dfrac{\partial H_{0}}{\partial \boldsymbol{\sigma}_j} \right)_{t}\right]dt  +d\boldsymbol{{w}}_i .
\end{equation}
Consistently with the above assumption, it is reasonable to look at the spin dynamics over a lower time resolution, because changes in the spins are negligible on the minimal scale $dt$. We can then consider a discrete time increment $\Delta t\gg \tau_{\mathrm m}$, which is much larger than the decay time of ${\cal R}$, but also much smaller than the time scale over which the spins vary significantly. The dynamical evolution of the spins on this coarse-grained timescale can be easily obtained by integrating Eq. \eqref{eq_of_motion_differential_3} between $t$ and $t+\Delta t$, and considering the spins  to be  constant in this integration interval. We get,
%
%
\begin{equation}\label{eq_of_motion_discrete}
\Delta \boldsymbol{\sigma}_i =  \lbrace H_0, \boldsymbol{\sigma}_i \rbrace \Delta t - \left[ \hbar^{-1} \left(\dfrac{\alpha^2}{\hbar K}\right)\tau_{\mathrm{m}}\sum_{j}     \Lambda_{ij}\left( \dfrac{\partial H_{0}}{\partial \boldsymbol{\sigma}_j} \right)_{t}\right]\Delta t  +\Delta \boldsymbol{{w}}_i ,
\end{equation}
where,
\begin{equation}
    \Delta \boldsymbol{{w}}_i = \int_{t}^{t+\Delta t} d\boldsymbol{{w}}_i
    \label{eq:discr-noise}
\end{equation}
Eq. \eqref{eq_of_motion_discrete} is now the lower time resolution version of Eq. \eqref{eq_of_motion_differential}. Memory is not present anymore in the relaxational term and $\Delta \boldsymbol{{w}}_i$  is the coarse-grained noise over scale $\Delta t$. From \eqref{eq:discr-noise} we find,
\begin{equation}\label{eq:correlation_noise_kernel_discrete_lower_resolution}
    \langle \Delta{w}^{\mu}_i(t_1) \Delta{w}^{\nu}_j(t_2) \rangle =  2k_BT \hbar^{-1} \left(\dfrac{\alpha^2}{\hbar K}\right) \Lambda_{ij}\delta_{\mu\nu} \int_{t_1}^{t_1+\Delta t}dt\int_{t_2}^{t_2+\Delta t} dt' {\cal R}(t-t') \ ,
\end{equation}
where $t_1$ and $t_2$ are multiples of the new time resolution $\Delta t$. Now, since $\Delta t$ is much larger than the decay time of ${\cal R}$, if $t_1\neq t_2$ Eq. \eqref{eq:correlation_noise_kernel_discrete_lower_resolution} gives 0; on the other hand, if $t_1=t_2$, all the times where the function is significantly different from 0 are inside the integration range, thus yielding the aforementioned $\tau_{\mathrm{m}}$. Therefore, we have
\begin{equation}\label{eq:correlation_noise_kernel_discrete_lower_resolution_final}
    \langle \Delta{w}^{\mu}_i(t_1) \Delta{w}^{\nu}_j(t_2) \rangle =  2k_BT \hbar^{-1} \left(\dfrac{\alpha^2}{\hbar K}\right) \tau_{\mathrm{m}} \Lambda_{ij}\delta_{\mu\nu}  \Delta t \;\delta_{t_1,t_2} \ .
\end{equation}
We then recover white uncorretaled noise (i.e. $\Delta{w}^{\mu}_i(t)$ is a Wiener process on scale $\Delta t$), together with the correct FD relation between its correlation and the relaxational coefficient.

Equation \eqref{eq_of_motion_discrete} is therefore Markovian and it has the very same structure on scale $\Delta t$  as the DLT equation used in the main text. This can be seen explicitly by discretizing Eq. \eqref{glip} of the main text on scale $\Delta t$, which becomes
\begin{equation}\label{eq_DLT_discrete_time_main}
    \Delta \boldsymbol{\sigma}_i =  \lbrace H_0, \boldsymbol{\sigma}_i \rbrace \Delta t - \left[ \hbar^{-1} \lambda\sum_{j}     \Lambda_{ij}\left( \dfrac{\partial H_{0}}{\partial \boldsymbol{\sigma}_j} \right)_{t}\right]\Delta t  +\Delta \boldsymbol{w}_i \ ,
\end{equation}
with
\begin{equation}
    \langle \Delta{w}^{\mu}_i(t_1) \Delta{w}^{\nu}_j(t_2) \rangle =  2k_BT \hbar^{-1} \lambda \Lambda_{ij}\delta_{\mu\nu}  \Delta t \;\delta_{t_1,t_2} \ .
    \label{eq:noise_dlt}
\end{equation}
By comparing these two equations with Eqs. \eqref{eq_of_motion_discrete} \eqref{eq:correlation_noise_kernel_discrete_lower_resolution_final} we can finally identify the relaxational coefficient $\lambda$ as,
\begin{equation}
    \lambda =\left(\dfrac{\alpha^2}{\hbar K}\right)\tau_{\mathrm{m}}  ,
\end{equation}
which is (finally) equation \eqref{bonny} of the main text.

\nocite{cavagna2023discrete}


\end{document}